\documentclass[journal]{IEEEtran}
\usepackage{makecell}
\usepackage{amssymb,amsthm,amsmath,bm}
\usepackage{multirow}
\usepackage{algorithm}
\usepackage{algorithmicx}
\usepackage{algpseudocode}
\usepackage{lipsum}
\usepackage{graphicx}
\usepackage{multicol}
\usepackage{float}
\usepackage{url}
\usepackage{color}
\usepackage{stfloats}
\usepackage{booktabs}
\usepackage{bm}
\usepackage{footnote}
\usepackage{graphicx}
\usepackage{rotating}
\usepackage{adjustbox}
\usepackage{verbatim}
\usepackage{float}

\ifCLASSINFOpdf
\else
\fi
\begin{document}
%
\title{Low-Light Maritime Image Enhancement with Regularized Illumination Optimization and\\Deep Noise Suppression}
%
%
%

\author{Yu Guo,
        Yuxu Lu, 
        Ryan Wen Liu,
        Meifang Yang,
        and Kwok Tai Chui
\thanks{This work was supported by the National Key R\&D Program of China (No.: 2018YFC0309602), and the National Natural Science Foundation of China (No.: 51609195). \textit{Yu Guo and Yuxu Lu are joint first authors.}}
\thanks{Y. Guo is with the School of Transportation, Wuhan University of Technology, Wuhan 430063, China (email: 272047@whut.edu.cn).}
\thanks{Y. Lu, R. W. Liu and M. Yang are with the Hubei Key Laboratory of Inland Shipping Technology, School of Navigation, Wuhan University of Technology, Wuhan 430063, China (e-mail: 1336201989@whut.edu.cn; wenliu@whut.edu.cn; youngmeifang@whut.edu.cn).}
\thanks{K. T. Chui is with the Department of Technology, School of Science and Technology, The Open University of Hong Kong, Hong Kong, China (e-mail: jktchui@ouhk.edu.hk).}
}

%
%

\markboth{Journal of \LaTeX\ Class Files,~Vol.~14, No.~8, August~2015}%
{Shell \MakeLowercase{\textit{et al.}}: Bare Demo of IEEEtran.cls for IEEE Journals}
%
%
%
%
%
%
\maketitle

\begin{abstract}
   Maritime images captured under low-light imaging condition easily suffer from low visibility and unexpected noise, leading to negative effects on maritime traffic supervision and management. To promote imaging performance, it is necessary to restore the important visual information from degraded low-light images. In this paper, we propose to enhance the low-light images through regularized illumination optimization and deep noise suppression. In particular, a hybrid regularized variational model, which combines L0-norm gradient sparsity prior with structure-aware regularization, is presented to refine the coarse illumination map originally estimated using Max-RGB. The adaptive gamma correction method is then introduced to adjust the refined illumination map. Based on the assumption of Retinex theory, a guided filter-based detail boosting method is introduced to optimize the reflection map. The adjusted illumination and optimized reflection maps are finally combined to generate the enhanced maritime images. To suppress the effect of unwanted noise on imaging performance, a deep learning-based blind denoising framework is further introduced to promote the visual quality of enhanced image. In particular, this framework is composed of two sub-networks, i.e., E-Net and D-Net adopted for noise level estimation and non-blind noise reduction, respectively. The main benefit of our image enhancement method is that it takes full advantage of the regularized illumination optimization and deep blind denoising. Comprehensive experiments have been conducted on both synthetic and realistic maritime images to compare our proposed method with several state-of-the-art imaging methods. Experimental results have illustrated its superior performance in terms of both quantitative and qualitative evaluations.
\end{abstract}

%
%
\IEEEpeerreviewmaketitle
%
\section{Introduction}
\label{sec:Introduction}
\subsection{Background and Related Work}
\IEEEPARstart{M}{aritime} images captured under low-light conditions often suffer from low contrast, poor visibility, and random noise. The captured low-light images easily fail to reflect valuable visual information, which will directly affect the effectiveness of many vision-based techniques, e.g., object detection \cite{FastRCNN, YOLO}, edge detection \cite{HED, RCF}, and visual navigation \cite{ZSVI, VNG}, etc. In practical applications, the low-light maritime images essentially suffer from the low-intensity luminance and noise corruption leading to the degradation of valuable visual information. To make low-light image enhancement more available, it is essential to effectively enhance the luminance contrast and suppress the unwanted noise. According to the important problems we focus in this work, we will briefly present the current progresses in low-light image enhancement and noise suppression.

\textbf{Low-Light Image Enhancement:} Traditional low-light enhancement methods can be roughly divided into histogram equalization-based methods, Retinex-based methods, and dehazing-based methods, etc. The classic histogram equalization method \cite{AHE} has been widely used in image contrast enhancement due to its advantages of time-domain processing, simple calculation, and easy implementation. To further promote the classic histogram equalization, several extended versions have been presented to produce more robust enhancement results. In the current literature, these methods can be categorized into global and local histogram equalization methods. The representative global histogram equalization methods, e.g., brightness preserving bi-histogram equalization (BBHE) \cite{BBHE}, minimum mean brightness error bi-histogram equalization (MMBEBHE) \cite{MMBEBHE}, and background brightness preserving histogram equalization (BBPHE) \cite{BBPHE}, etc., were proposed to enhance the flexibility of histogram equalization. These methods established the dual histogram equalization strategy by decomposing the original histogram into two histograms to enhance low-light images. Besides, brightness preserving histogram equalization with maximum entropy (BPHEME) \cite{BPHEME} proposed a histogram variation technique, which combined image processing theory with optimization theory and functional analysis. BPHEME could maintain luminance and preserve local details better compared with MMBEBHE. Many local histogram equalization methods were proposed \cite{POSHE, ABMHE, CMBFHE, CNSLHE} since the histogram equalization with a single conversion function was difficult to enhance contrast in the dark regions. They thus performed local histogram equalization methods to enhance the local details. However, these methods often suffer from noise residue and over-enhancement in practical applications.

Based on the assumption of Retinex theory \cite{Retinex}, the observed image can be decomposed into the illumination and reflection maps. The reflection map contains the intrinsic color and important geometrical structures. In contrast, only the illumination map, which is smoothly varying, contains the luminance information. Early Retinex-based attempts \cite{SSR, MSR, MSRCR} tended to adopt the Gaussian filter to estimate the smooth illumination maps and directly consider the reflection maps as the final enhanced images. The enhanced results thus have plentiful details and high-intensity illumination. However, they often ignored the influences of illumination maps on image enhancement leading to negative effects in several different ways, such as over-enhancement and unnaturalness. In order to generate more natural-looking images, it is particularly important to optimally refine the illumination map. Kimmel \textit{et al.} \cite{AVF} first proposed a variational framework to estimate the smooth illumination map. However, the estimation of reflection map is lacking in the proposed framework leading to limiting the improvement of image quality. To further improve imaging performance, a low-light image enhancement algorithm for non-uniform illumination images \cite{NPEA} has been proposed to restore the details and preserve the naturalness. In \cite{BCP}, a bright channel prior (BCP)-based image restoration method was presented to obtain a satisfactory illumination map. In particular, BCP could eliminate the black halo and suppress the color distortion better. Fu \textit{et al.} \cite{ANRBA} proposed a novel Retinex-based image enhancement method with illumination adjustment. The proposed method performs well in preserving significant edges in reflection and properly adjusting illumination. The naturalness preservation and detail enhancement could be correspondingly generated in the enhanced images. To further promote image quality, a weighted variational model for simultaneous reflection and illumination estimation (SRIE) \cite{SRIE} was presented. SRIE is able to preserve the estimated reflection with more details and suppress random noise to some extent. The quality of enhanced images could be improved accordingly.

From the statistical point of view, the inverted low-light images are visually similar to the degraded images captured under hazy weather conditions. Several methods \cite{DeHZ, NVE, BDCE} have been proposed based on the assumption of dark channel prior (DCP) \cite{DCP}, which was originally presented to perform image dehazing. In particular, Dehazing-based methods first inverted the low-light images and then adopted the improved dark channel prior method to deal with the inverted images. Furthermore, many strategies, e.g., local smoothing, Gaussian pyramid operators, block-matching and 3D filtering (BM3D) \cite{BM3D}, etc, were employed to improve the image quality. Finally, the enhanced results could be obtained by inverting the dehazed images again. Dehazing-based enhancement methods can effectively improve low-light intensities, but they often fail to further enhance visual quality due to the lack of theoretical basis.

With the rapid development of deep learning, the conventional neural network (CNN) \cite{Deep L} has been widely applied in the fields of image processing and computer vision. The low-light image enhancement has gained great achievement by taking full advantage of deep learning. For example, Lore \textit{et al.} \cite{LLNet} proposed a deep autoencoder-based learning approach (LLNet), which could identify signal features from low-light images and adaptively improve the luminance without over-amplifying the lighter regions. Chen \textit{et al.} \cite{LTS} presented a fully convolutional network structure to process low-light images with end-to-end training mode. They fully considered the influences of long and short exposures on imaging under low-light conditions to construct the datasets. Therefore, the better enhancement results on realistic low-light images could be produced accordingly. Furthermore, Hui \textit{et al.} \cite{PPCN} proposed a perception-preserving convolution network (PPCN) to learn the mapping between ordinary photos and DSLR-quality images. The Retinex-Net \cite{Retinex-Net} was different from other end-to-end networks, which was composed of a Decom-Net and an Enhance-Net. In particular, the Decom-Net was employed to estimate the illumination map, and the Enhance-Net was employed to adjust the illumination map. It is worth mentioning that the deep learning-based image enhancement methods are strongly dependent on the volume and diversity of training datasets. It is often difficult to produce satisfactory image quality if the training datasets do not contain the geometrical features existed in images to be enhanced. In this work, we will only adopt the deep learning as a post-processing step to further enhance visual quality.

\textbf{Noise Suppression:} The representative traditional denoising methods, e.g., adaptive variational method \cite{LiuShiMRI2014}, patch-based nonlocal means (NLM) \cite{NLM}, weighted nuclear norm minimization (WNNM) \cite{WNNM}, and BM3D \cite{BM3D}, can effectively eliminate random noise. However, these methods essentially suffer from two main drawbacks: (1) time-consuming and often fail to reduce spatially variant noise; (2) difficult to achieve satisfactory denoising performance when the noise level is unknown. The latest generation of deep learning technology has achieved remarkable successes in image denoising. For example, the denoising convolution neural network (DnCNN) \cite{DnCNN} was originally proposed to suppress the white Gaussian noise. DnCNN was committed to obtaining the mapping function between the input degraded image and the output noise-free image through the residual learning strategy \cite{Resiadual L}. However, the realistic noise is commonly non-uniform Gaussian distributed, which may be changed by the spatial domain in practice. To handle this problem, Zhang \textit{et al.} \cite{FFDNet} proposed a fast and flexible solution for CNN-based image denoising (termed FFDNet), which could deal with the spatially variant noise with different levels. More recently, Guo \textit{et al.} \cite{CBDNet} presented a convolutional blind denoising network (CBDNet) to boost the blind denoising performance and improve the generalization ability of network. In particular, the CBDNet was composed of a $5$-layer fully convolutional network and a $16$-layer U-Net \cite{UNet}. The $5$-layer fully convolutional network was used to estimate the noise level, and the 16-layer U-Net was used to suppress the random noise. Experiments have shown that CBDNet is capable of generating satisfactory denoising performance is the case of unknown noise level. In this work, a blind deep denoising strategy will be adopted as a post-processing step to optimize the enhanced images to improve image quality.
\subsection{Motivation and Contributions}
It is well known that Retinex theory \cite{Retinex} is a crucial assumption in the fields of image processing and computer vision. Many deep learning-based methods have been proposed based on this assumption \cite{Retinex-Net, RDGAN}. To make image enhancement more satisfactory, we propose a two-step framework for low-light image enhancement based on the Retinex theory, which benefits from the regularized illumination optimization and deep blind denoising. In particular, the Retinex theory considers that both illumination and reflection maps jointly constitute the observed color image. In the current literature, many regularized variational models \cite{AVF, NPEA, BCP, ANRBA, SRIE, ADPV} have been currently adopted to estimate the smoothed illumination maps. These estimation methods, however, inevitably smooth the edge structures leading to visual quality degradation. According to the Retinex theory, the estimated illumination map should retain the significant edge structures while smoothing the textural details. To achieve this requirement, we propose a hybrid regularized variational model, which combines the L2-norm data-fidelity term, L0-norm gradient sparsity prior \cite{L0}, and relative total variation (RTV) regularizer \cite{RTV}. The L2-norm data-fidelity term is used to suppress the generation of outliers. The L0-norm and RTV constraints can retain the important geometrical structures and smooth the textural details. To guarantee a stable solution, the resulting non-smooth optimization problem will be handled using an effective numerical algorithm \cite{ADMM}. Meanwhile, random noise existed in original low-light images could lead to visual quality degradation. Many existing methods \cite{Retinex-Net, LIME} proposed to denoise the estimated reflection map to enhance image quality. However, from an imaging point of view, the reflection map is significantly different from the observed original image. The denoising of reflection map could cause severe color distortions in enhanced images leading to degraded visual quality. In this work, we will introduce a blind denoising network to denoise the final enhanced images to further enhance imaging quality.

In particular, we will propose to incorporate both L0-norm regularizer and RTV into a regularized variational model to guarantee more robust illumination refinement. From a theoretical point of view, the refined illumination maps, only using L0-norm gradient minimization, easily suffer from various artifacts, e.g., over-sharpening effects and under-filtering of high-amplitude textures, etc \cite{L0}. RTV adopted in LIME \cite{LIME} is effective in removing texture, but sometimes blurs the major edges. The combination of L0-norm and RTV could effectively overcome these limitations and improve the illumination refinement results. The quality of enhanced images can be significantly improved accordingly. To reduce the effect of random noise on image enhancement, the deep learning method will be introduced to blindly remove the unwanted noise with unknown noise level. However, how to use this deep learning strategy to enhance image quality is still an important problem. There are three strategies considered to incorporate deep learning into low-light image enhancement in this work. We will discuss the influences of these strategies on image enhancement and select the best one in our low-light image enhancement framework.

In the current literature, low-light image enhancement mainly focuses on indoor screens or outdoor natural images. Few studies have been conducted on low-light maritime images. Meanwhile, we find that almost no low-light image enhancement methods can suppress random noise blindly. Compared with previous studies, the major contributions of our work can be summarized as follows
\begin{enumerate}
	\item A unified image enhancement framework, which involves illumination refinement, reflection optimization, and deep blind denoising, is developed to improve image quality under low-light conditions. It has the capacity of enhancing visual quality, blindly reducing random noise, and suppressing unwanted artifacts.
	\item A hybrid regularized variational model, which combines L0-norm gradient sparsity prior with structure-aware regularization, is proposed to refine the illumination map. The proposed model has the capacity of preserving the significant structures and removing the textural details during illumination optimization.
	\item The quality of enhanced image is further promoted using a blind denoising framework. This framework is composed of two sub-networks, i.e., E-Net and D-Net adopted for noise level estimation and non-blind noise reduction, respectively. The introduced blind denoising framework is able to effectively reduce the unwanted noise under poor imaging conditions. 
	\item Extensive experiments on both synthetic and realistic low-light images have demonstrated the superior imaging performance of our proposed enhancement method. Our method is capable of enhancing the low-light images and suppressing the unsatisfactory artifacts.
\end{enumerate}

The main benefit of our proposed method is that it takes full advantage of the regularized illumination optimization and deep blind denoising. Thus it can effectively enhance low-light images, suppress unwanted random noise and preserve fine structural details in practice.
\subsection{Organization}
The remainder of this paper is divided into the following sections. Section \ref{sec:PF} mainly describes the problem formulation related to low-light image enhancement. In Section \ref{sec:IOUNCVM}, a hybrid regularized variational model is proposed to refine the coarse illumination map estimated using Max-RGB. Section \ref{sec:PLLIEF} is devoted to generating enhanced images, which are further visually promoted through the blind denoising framework. Experiments on both synthetic and realistic maritime images are implemented in Section \ref{sec:ERD}. Finally, we conclude our main contributions in Section \ref{sec:C}.
\section{Problem Formulation}
\label{sec:PF}
\begin{figure}[t]
	\centering
	\includegraphics[width=1\linewidth]{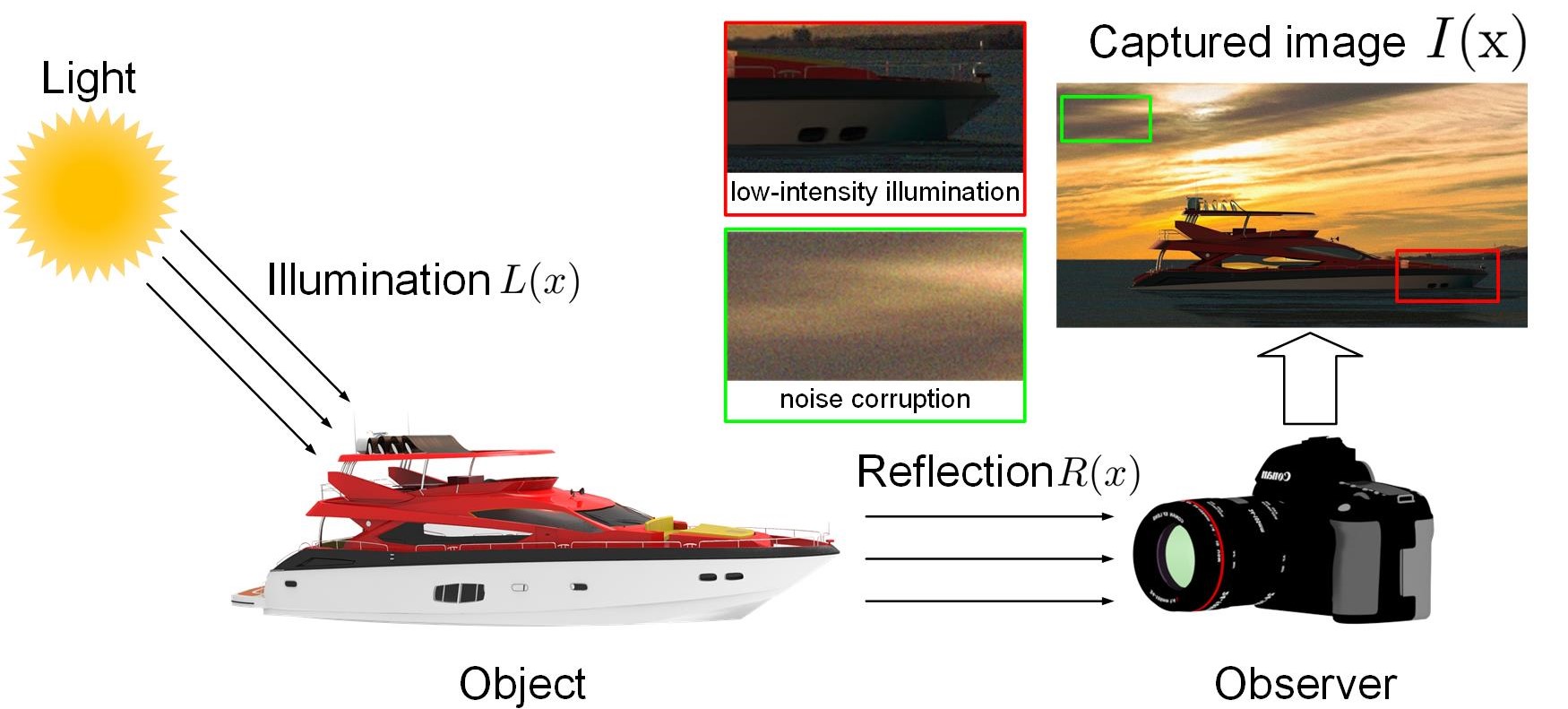}
	\caption{The principle of Retinex theory.}
	\label{Retinex}
\end{figure}

\begin{figure*}[t]
	\centering
	\includegraphics[width=0.75\linewidth]{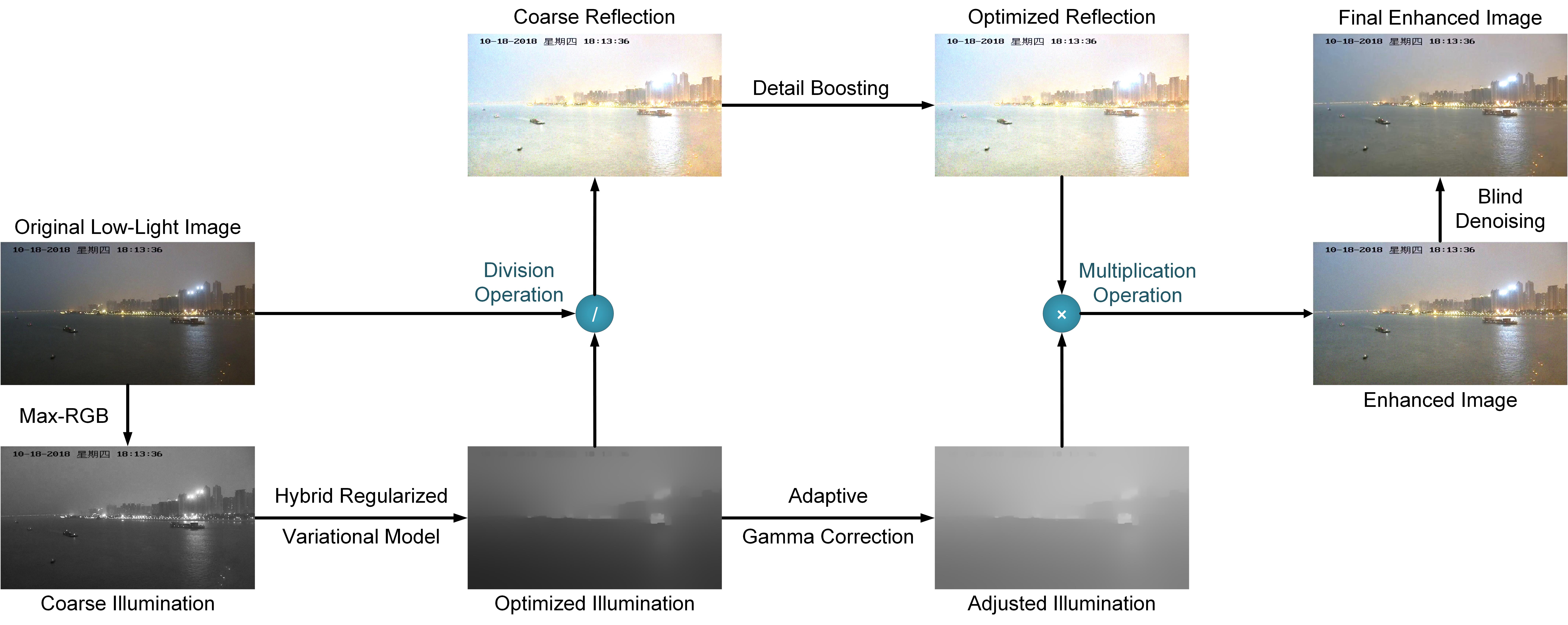}
	\caption{Flowchart of our proposed method for enhancing low-light maritime images.}
	\label{Fig2}
\end{figure*}

In the current literature \cite{YangITSC2019}, low-light enhancement methods are mainly proposed based on the assumption of Retinex theory. The principle of Retinex theory can be visually illustrated in Fig. \ref{Retinex}. The captured low-light image $I$ can be decomposed as follows 
\begin{equation}\label{Eq:Retinex}
{I(\mathrm{x}) = R(\mathrm{x}) \circ L(\mathrm{x})},
\end{equation}
where $\mathrm{x} \in \Omega$ denotes the pixel with $\Omega$ being the image domain, $L$ and $R$ represent the illumination and reflection maps, respectively. The operator $\circ$ is an element-wise multiplication operator. In this work, we assume that all three channels (i.e., RGB) of color images have the same illumination. It can be seen from Fig. \ref{Retinex} that only the illumination map $L$ contains the luminance information affected by light. In contrast, the reflection map $R$ essentially contains textural details and random noise affected by the object and the shooting process. It is worth noting that two partially magnified views of the captured image under low-light imaging conditions in Fig. \ref{Retinex} often suffer from low-intensity illumination and noise corruption. Therefore, to enhance imaging quality, it is necessary to estimate and adjust the illumination map, boost the textural details existed in the reflection map, and suppress the undesirable noise.

Retinex-based image enhancement methods could be roughly divided into two categories. The first type considers the reflection map as the final enhanced image. In particular, these methods directly remove the illumination map and optimize the reflection map. Another type tends to recombine the illumination and reflection maps to restore the low-light image. To improve imaging performance, they propose to optimize both reflection and illumination maps. The first type is not only susceptible to distortion but also easily causes the over-enhancement problem due to the direct removal of illumination map. Recent studies have shown that it is more reasonable to yield satisfactory results by jointly optimizing both illumination and reflection maps.

The flowchart of our image enhancement method is summarized in Fig. \ref{Fig2}. In the first step, a hybrid regularized variational model is proposed to refine the coarse illumination map originally estimated using Max-RGB. In the second step, an adaptive gamma correction method and a guided filter-based detail boosting method are adopted to optimize the reflection map. The refined illumination and optimized reflection maps are combined to generate the final enhanced images. As a post-processing step, the blind denoising framework is introduced to reduce the unwanted noise to further improve visual image quality.
\section{Regularized Illumination Optimization}
\label{sec:IOUNCVM}
The performance of Retinex-based low-light enhancement method depends, to a great extent, upon the estimation of illumination map. To obtain a satisfactory illumination map, we first adopt the Max-RGB method to estimate the coarse illumination map. The hybrid regularized variational model is then proposed to further generate the refined illumination map leading to image quality improvement.
\subsection{Coarse Illumination Map Using Max-RGB}
\label{sub-sec:MR}
The popular Max-RGB method, widely adopted for coarse illumination estimation, is essentially related to the dark channel prior (DCP) \cite{DCP}. Since the inverted low-light image $(1 - I)$ looks similar to the hazy image, we can transform the atmospheric scattering model (\ref{Eq:Retinex}) as follows
\begin{equation}\label{Eq:Max-RGB1}
{1-I(\mathrm{x}) = (1-R(\mathrm{x})) \circ \tilde{L}(\mathrm{x}) + A (1 - \tilde{L}(\mathrm{x}))},
\end{equation}
with $A$ being the global atmospheric light. Note that DCP has been widely used to estimate the transmission map for image dehazing. Therefore, we still tend to adopt the DCP to initially estimate the illumination map, i.e.,
%
\begin{equation}\label{Eq:Max-RGB2}
\begin{aligned}
\tilde{L}(\mathrm{x}) &\approx 1-\min_{c \in \{R,G,B\}}\dfrac{1-I^{c}(\mathrm{x})}{A} \\
&=1 - \dfrac{1}{A}+\max_{c \in \{R,G,B\}}\dfrac{I^{c}(\mathrm{x})}{A},
\end{aligned}
\end{equation}
where $\tilde{L}$ is the coarse illumination map, and $I^{c}$ represents the single-channel image of the RGB image $I$ in channel $c \in \{R,G,B\}$. Please refer to \cite{DCP} for more details on DCP-based transmission map estimation. We consider that all three channels (i.e., RGB) of color images have the same illumination. The intensities in low-light images will significantly become large in hazy images after inversion. From a statistical point of view, the global atmospheric light $A$ is close to $1$. In this work, we roughly set $A=1$ since the inverted low-light image is similar to the hazy image. Eq. (\ref{Eq:Max-RGB2}) can thus be rewritten as follows
\begin{equation}\label{Eq:Max-RGB3}
\tilde{L}(\mathrm{x}) \approx \max_{c \in \{R,G,B\}}I^{c}(\mathrm{x}),
\end{equation}
which will be directly adopted to estimate the coarse illumination map $\tilde{L}$ in our experiments.
\subsection{Refined Illumination Estimation Using Hybrid Regularized Variational Model}
\label{sub-sec:NC}

To obtain the satisfactory enhancement results, the estimated illumination should smooth the texture details while preserving the main geometrical structures. However, the illumination map estimated by Max-RGB obviously failed to achieve this claim. It is thus necessary to further optimize the illumination map. In this paper, a nonsmooth nonconvex regularized optimization model is proposed to refine the coarse illumination map, i.e.,
\begin{equation}\label{eq:refinedmodel}
\min_{\hat{L}} \left\{ \frac{1}{2} \big\| \hat{L} - \tilde{L} \big\|_{2}^2 + \lambda_1 \mathcal{L}_0 ( \hat{L} ) + \lambda_2 \mathcal{R} ( \hat{L} ) \right\}
\end{equation}
where $\lambda_1$ and $\lambda_2$ are positive regularization parameters, $\tilde{L}$ and $\hat{L}$ denote the coarse and refined illumination maps.

The first term in Eq. (\ref{eq:refinedmodel}) is named the squared L2-norm fidelity term, which can guarantee the solution accords with the degradation process and suppress the potential outliers. The second term is the L0-norm prior which can smooth the low-amplitude structures and enhance the salient edges, which can be defined as follows
\begin{equation}\label{eq:L0norm}
\mathcal{L}_0 ( \hat{L} ) = \big\| \nabla_{h} \hat{L} \big\|_{0} + \big\| \nabla_{v} \hat{L}  \big\|_{0},
\end{equation}
where $\nabla_{h}$ and $\nabla_{v}$, respectively, denote the first-order derivative filters in horizontal and vertical directions. The third term $\mathcal{R} ( \cdot )$ in Eq. (\ref{eq:refinedmodel}) is selected as the RTV regularizer \cite{RTV}, which can assist in preserving meaningful structures during illumination optimization. The theoretical definition of RTV for one pixel $\mathrm{x} \in \Omega$ is given by
\begin{equation}\label{eq:DefRTV}
\mathcal{R} ( \hat{L} (\mathrm{x}) ) = \frac{\mathcal{D}_{h} (\mathrm{x})}{ \mathcal{L}_{h} (\mathrm{x}) + \varepsilon} + \frac{\mathcal{D}_{v} (\mathrm{x})}{ \mathcal{L}_{v} (\mathrm{x}) + \varepsilon},
\end{equation}
where $\varepsilon > 0$ is a constant to avoid division by zero, $\mathcal{D}_{*} ( \cdot )$ and $\mathcal{L}_{*} ( \cdot )$ with $* \in \{ h, v \}$, respectively, denote the windowed total variation and windowed inherent variation, i.e.,
\begin{equation}\label{eq:WTV}
\mathcal{D}_{*} ( \mathrm{x} ) = \sum\nolimits_{\mathrm{y} \in \bar{\Omega} (\mathrm{x})} g (\mathrm{x}, \mathrm{y}) \cdot \big| (\nabla_{*} \hat{L}) (\mathrm{y}) \big|,
\end{equation}
\begin{equation}\label{eq:WTV}
\mathcal{L}_{*} ( \mathrm{x} ) = \big| \sum\nolimits_{\mathrm{y} \in \bar{\Omega} (\mathrm{x})} g (\mathrm{x}, \mathrm{y}) \cdot (\nabla_{*} \hat{L}) (\mathrm{y}) \big|,
\end{equation}
with $\bar{\Omega} (\mathrm{x})$ denoting the region centered at pixel $\mathrm{x} \in \Omega$, and the weighting function $g (\mathrm{x}, \mathrm{y})$ being given by
\begin{equation}\label{eq:WeightingFun}
g (\mathrm{x}, \mathrm{y}) \propto \exp \left( - \frac{ (\mathrm{x}_h - \mathrm{y}_h )^2 + (\mathrm{x}_v - \mathrm{y}_v )^2 }{2 \sigma^2} \right),
\end{equation}
where $\mathrm{x} = (\mathrm{x}_h, \mathrm{x}_v ) \in \Omega$ and $\mathrm{y} = (\mathrm{y}_h, \mathrm{y}_v ) \in \Omega$, $\sigma$ is the standard deviation which controls the spatial scale of the region $\bar{\Omega} ( \cdot )$. The combination of L0-norm and RTV has the capacity of suppressing significant structures and removing textural details in the optimized illumination map. The quality of enhanced images could be improved accordingly. However, due to the nonsmooth and nonconvex natures of regularizers in Eq. (\ref{eq:refinedmodel}), it is intractable to effectively handle Eq. (\ref{eq:refinedmodel}) using simple numerical method \cite{LuDuanMMAS2016}.
\begin{figure*}[t]
	\centering
	\includegraphics[width=1\linewidth]{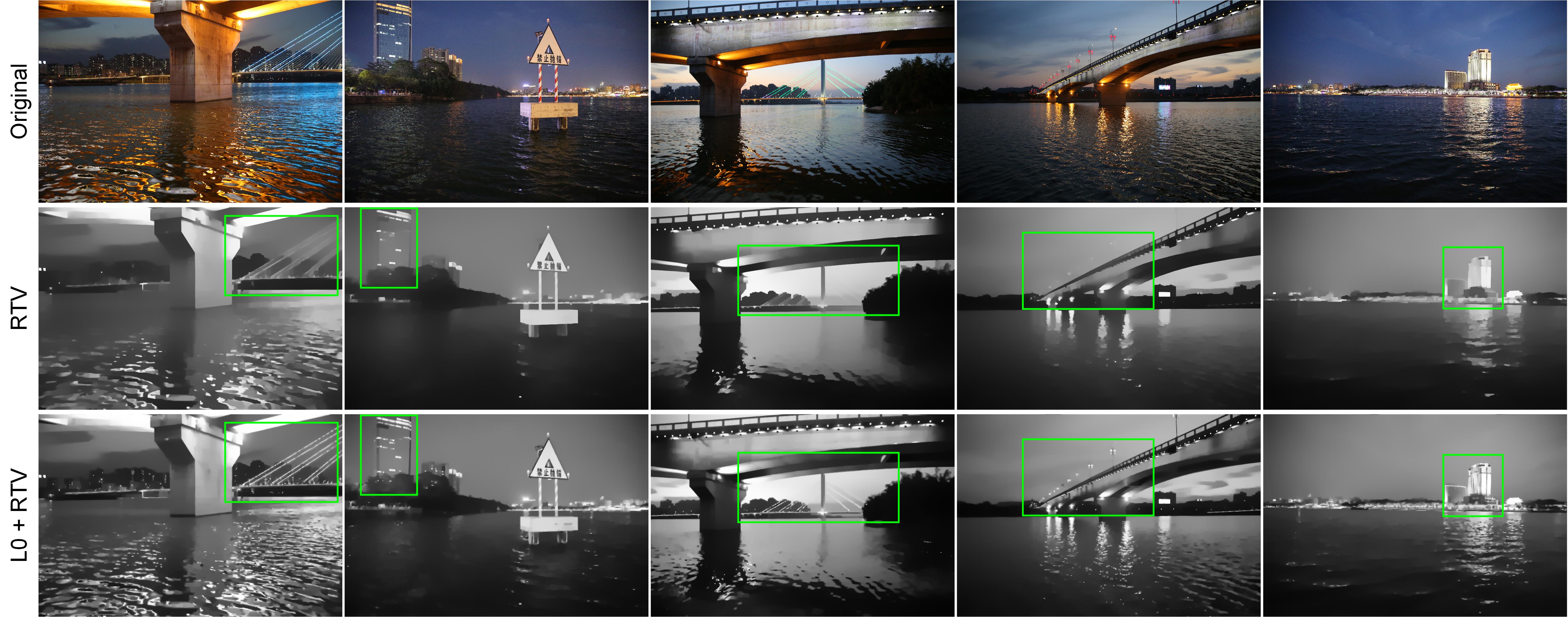}
	\caption{Visual displays of refined illumination maps for different low-light images. From top to bottom: original low-light images, refined illumination maps generated by only RTV, and the combination of L0-norm and RTV.}
	\label{fig:illuminationmap}
\end{figure*}
\subsubsection{Two-Step Optimization Approach}
To achieve stable solution, we introduce two intermediate variables $U_h = \nabla_{h} \hat{L}$ and $U_v = \nabla_{v} \hat{L}$ and transform the unconstrained minimization problem (\ref{eq:refinedmodel}) into the following constrained version
\begin{align}\label{Eq:constrainedproblem}
&\min_{U_h, U_v, \hat{L}} \Big\{ \frac{1}{2} \big\| \hat{L} - \tilde{L} \big\|_{2}^2 + \lambda_1 \big( \big\| U_h \big\|_{0} + \big\| U_v  \big\|_{0} \big) + \lambda_2 \mathcal{R} ( \hat{L} ) \Big\}\nonumber \\
&~~~~\mathrm{s.t.}~~~U_h = \nabla_{h} \hat{L}, U_v = \nabla_{v} \hat{L},
\end{align}
whose equivalent version can be obtained as follows
\begin{align}
\min_{U_h, U_v, \hat{L}} & \Big\{ \frac{1}{2} \big\| \hat{L} - \tilde{L} \big\|_{2}^2 + \lambda_1 \big( \big\| U_h \big\|_{0} + \big\| U_v  \big\|_{0} \big) + \lambda_2 \mathcal{R} ( \hat{L} ) \nonumber \\
& + \frac{\beta_1}{2} \big\| U_h - \nabla_{h} \hat{L} \big\|_{2}^2 + \frac{\beta_2}{2} \big\| U_v - \nabla_{v} \hat{L} \big\|_{2}^2 \Big\}, \label{Eq:unproblem2}
\end{align}
where $\beta_1$ and $\beta_2$ are positive constant parameters. If $\beta_1 \rightarrow \infty$ and $\beta_2 \rightarrow \infty$, the solutions in Eq. (\ref{Eq:unproblem2}) will be equivalent to the solutions in Eq. (\ref{Eq:constrainedproblem}). In this work, the unconstrained optimization problem (\ref{Eq:unproblem2}) will be effectively solved using a two-step optimization approach \cite{LiuShiMP2015} which iteratively minimizes with respect to $(U_h, U_v)$ and $\hat{L}$ separately. In particular, the two steps we will perform are given by
\begin{align}
\mathrm{Step~1}:~\min_{U_h, U_v} & \Big\{ \lambda_1 \big\| U_h \big\|_{0} + \lambda_1 \big\| U_v  \big\|_{0} \label{Eq:Step1} \\
&+ \frac{\beta_1}{2} \big\| U_h - \nabla_{h} \hat{L} \big\|_{2}^2 + \frac{\beta_2}{2} \big\| U_v - \nabla_{v} \hat{L} \big\|_{2}^2 \Big\}, \nonumber
\end{align}
\begin{align}
\mathrm{Step~2}:~\min_{\hat{L}} & \Big\{ \frac{1}{2} \big\| \hat{L} - \tilde{L} \big\|_{2}^2 + \lambda_2 \mathcal{R} ( \hat{L} ) \label{Eq:Step2} \\
& + \frac{\beta_1}{2} \big\| U_h - \nabla_{h} \hat{L} \big\|_{2}^2 + \frac{\beta_2}{2} \big\| U_v - \nabla_{v} \hat{L} \big\|_{2}^2 \Big\}. \nonumber
\end{align}

In the Step 1 of our two-step optimization approach, the minimizations of $U_h$ and $U_v$ are essentially related to the standard L0-norm optimization problem. Therefore, the solutions $U_h$ and $U_v$ could be easily obtained using the element-wise hard thresholding operator, i.e.,
\begin{equation}\label{eq:solutionsUhv}
U_h = \mathcal{H}_{\lambda_1, \beta_1} \big( \nabla_{h} \hat{L} \big),~\mathrm{and}~U_v = \mathcal{H}_{\lambda_1, \beta_2} \big( \nabla_{v} \hat{L} \big),
\end{equation}
where the definition of $\mathcal{H}_{a, b} ( \cdot )$ is given by
\begin{align*}
\mathcal{H}_{a,b} \left( {s} \right)  &=
\begin{cases}
0,& \mathrm{if} \left| {s} \right| < \sqrt{{2a}/{b}},\\
{s}, & \mathrm{otherwise}.
\end{cases}
\end{align*}
with both $a$ and $b$ being intermediate variables.

The $\hat{L}$-subproblem in Step 2 is essentially a least-squares optimization problem constrained by RTV regularizer. Let $\mathcal{F}(\hat{L}) = \frac{1}{2} \big\| \hat{L} - \tilde{L} \big\|_{2}^2 + \frac{\beta_1}{2} \big\| U_h - \nabla_{h} \hat{L} \big\|_{2}^2 + \frac{\beta_2}{2} \big\| U_v - \nabla_{v} \hat{L} \big\|_{2}^2$ which is a smooth convex function. To achieve a numerically stable solution, the proximal forward-backward splitting (PFBS) framework \cite{CombettesMMS2005} will be introduced to solve the $\hat{L}$-subproblem in Step 2. In particular, the PFBS-based iterative thresholding algorithm for effectively handling Eq. (\ref{Eq:Step2}) is given by
\begin{equation}
\small
\begin{cases}
\begin{split}
\bar{L} &\leftarrow \hat{L} - t \nabla \mathcal{F}(\hat{L}) \label{SubproblemLbar} \\
\hat{L} &\leftarrow \min\nolimits_{\hat{L}} \Big\{ \big\| \hat{L} - \bar{L} \big\|_{2}^2 + \bar{\lambda}_2 \mathcal{R} ( \hat{L} ) \Big\} \\
\end{split}
\end{cases}
\end{equation}
with $\bar{\lambda}_2 = 2 t \lambda_2$ and $\nabla \mathcal{F}(\hat{L}) = \hat{L} - \tilde{L} + \beta_1 \nabla_{h}^T (\nabla_{h} \hat{L} - U_h)+ \beta_2 \nabla_{v}^T (\nabla_{v} \hat{L} - U_v)$. It is obvious that $\min_{\hat{L}} \big\{ \big\| \hat{L} - \bar{L} \big\|_{2}^2 + \bar{\lambda}_2 \mathcal{R} ( \hat{L} ) \big\}$ is essentially related to image filtering regularized by RTV proposed in \cite{RTV}. It is able to decompose the RTV regularizer into a nonlinear term (i.e., essentially weighting parameters) and a quadratic term. The approximated nonlinear optimization problem could be decomposed into a set of subproblems which are much easier to solve effectively. We refer the interested reader to \cite{RTV} for more details on numerical solution for RTV-regularized image filtering. We alternatively implement the iterative threshold algorithm (\ref{SubproblemLbar}) until the obtained solution converges to the optimal one. The advantage of our combination of L0-norm and RTV on illumination refinement is confirmed by the visual comparisons in Fig. \ref{fig:illuminationmap}. It is obvious that the refined illumination maps, only using RTV regularization, easily suffer from the loss of prominent structures. In contrast, our combined version has the capacity of preserving significant structures and removing textural details during illumination optimization. The quality of final enhanced image can then be improved accordingly.
\subsubsection{Computational Complexity and Convergence Analysis}
The proposed two-step optimization algorithm is in principle simple as it is intuitive. Let $k$ denote the total number of outer iterations, and $M \times N$ be the image size. In our imaging experiments, we directly select the total number of inner iteration as $1$ in Eq. (\ref{SubproblemLbar}). The computational cost mainly involves two parts, i.e., numerical solutions in Step 1 and Step 2. The hard threshold operators for both $U_h$ and $U_v$ can be easily performed with $\mathcal{O} (2MN)$ operations in Step 1. It is more complicated to analyze the time complexity of PFBS in Step 2. The computational bottleneck is due to the solution of RTV-regularized least-squares problem in Eq. (\ref{SubproblemLbar}). Inspired by the work \cite{RTV}, it becomes easy to analyze the computational complexity of our PFBS, i.e., $\mathcal{O} ( (\sigma + 1) MN)$. For illumination refinement, the total computational complexity of our two-step optimization algorithm can be theoretically obtained as $\mathcal{O} ( k (\sigma + 3) MN)$. The comparisons of running time for different image enhancement methods will be detailedly illustrated in Table \ref{table7}.

Note that we proposed a two-step optimization algorithm which decomposed the original minimization problem (\ref{eq:refinedmodel}) into two simple subproblems. Each subproblem could be effectively solved using existing numerical method. This optimization strategy has been successfully introduced to handing variational image restoration \cite{HuangSIAM2013}. The corresponding convergence has already been proved theoretically\footnote{Please refer to Ref. \cite{HuangSIAM2013} for more detailed information.}. Analogous to Ref. \cite{HuangSIAM2013}, the closed-form solutions of $(U_h, U_v)$-subproblems in Step 1 can be exactly obtained using the hard thresholding operators (\ref{eq:solutionsUhv}). The $\hat{L}$-subproblem in Step 2 is essentially a least-squares optimization problem constrained by a convex regularizer. Therefore, the convergence of the corresponding numerical solution can be guaranteed since the convergence of PFBS has been established to solve generic convex optimization problems \cite{CombettesMMS2005,AttouchMP2013}. We remark that it is tractable to obtain the closed-form solutions of $(U_h, U_v)$-subproblems in Step 1. Solution of Eq. (\ref{SubproblemLbar}) is also a global minimizer of $\hat{L}$-subproblem in Step 2. Based on the Opial theorem \cite{OpialBAMS1967}, the iterative sequence $\{ \hat{L} \}$ in Eq. (\ref{SubproblemLbar}) converges to a fixed point of the $\hat{L}$-subproblem in Step 2, i.e., a minimizer of minimization problem (\ref{eq:refinedmodel}). The convergence of our two-step optimization algorithm can thus be guaranteed accordingly.
\section{Low-Light Image Enhancement with Deep Noise Suppression}
\label{sec:PLLIEF}
This section is dedicated to generating the final enhanced image by combining refined illumination and optimized reflection maps. The optimized reflection map is obtained using the adaptive gamma correction and detail boosting methods. The deep learning method is further introduced to blindly remove the potential (unwanted) noise existed in enhanced image.
\subsection{Illumination Adjustment}
\label{sub-sec:IA}
It is well known that low-light images have low-intensity illumination in the dark regions. The illumination map essentially contains aplenty luminance information which is tightly related to visual image quality. It is thus necessary to adjust the illumination maps to generate satisfactory enhancement results. In the literature \cite{SRIE, LIME}, gamma correction has been widely employed to adjust the illumination map. The adjusted illumination map $\hat{L}_G$ is accordingly obtained using the gamma correction, i.e.,
\begin{equation}\label{Eq:Gamma Correction1}
\hat{L}_G (\mathrm{x}) = \hat{L}^{\frac{1}{\gamma}} (\mathrm{x}),
\end{equation}
where $\gamma$ is a coefficient larger than $1$. If $\gamma$ becomes larger, the enhancement effect will be more obvious. However, this method fails to adaptively adjust the illumination map, i.e., the illumination map is sensitive to the constant coefficient $\gamma$. To overcome this limitation, we propose to adaptively adjust $\gamma$ through calculating the average pixel value of the illumination map, i.e.,
\begin{equation}\label{Eq:Gamma Correction2}
{\gamma (\mathrm{x}) = \left\{
	\begin{aligned}
	&\gamma_0 \frac{\log \mu_{\hat{L}} (\mathrm{x}) } {\log0.5} &, \mu_{\hat{L}}(\mathrm{x}) \leq 0.5, \\
	&\gamma_0 &, \mu_{\hat{L}}(\mathrm{x}) > 0.5,
	\end{aligned}
	\right.}
\end{equation}
where $\mu_{\hat{L}} (\mathrm{x}) $ denotes the local mean value of $\hat{L}$ within the local region around $\mathrm{x} \in \Omega$, $\gamma_0$ is a pre-selected coefficient related to the enhancement intensity. 
\begin{figure*}[t]
	\centering
	\includegraphics[width=1\linewidth]{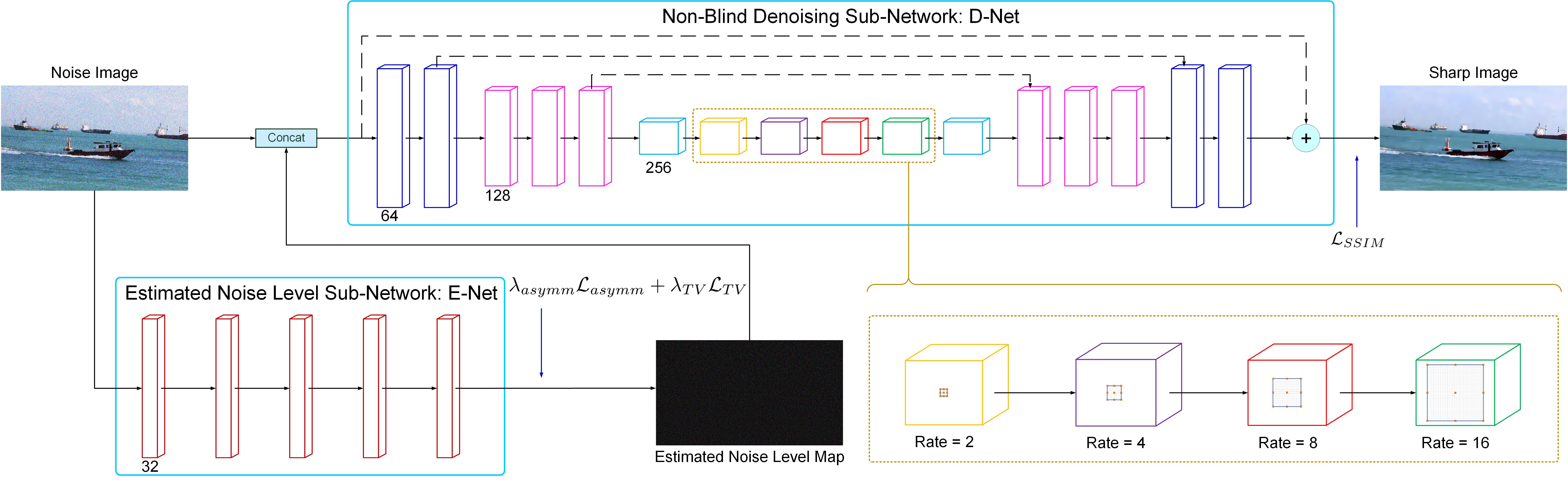}
	\caption{The architecture of blind denoising network used in this work.}
	\label{Blind}
\end{figure*}
\begin{figure*}[t]
	\centering
	\includegraphics[width=1.0\linewidth]{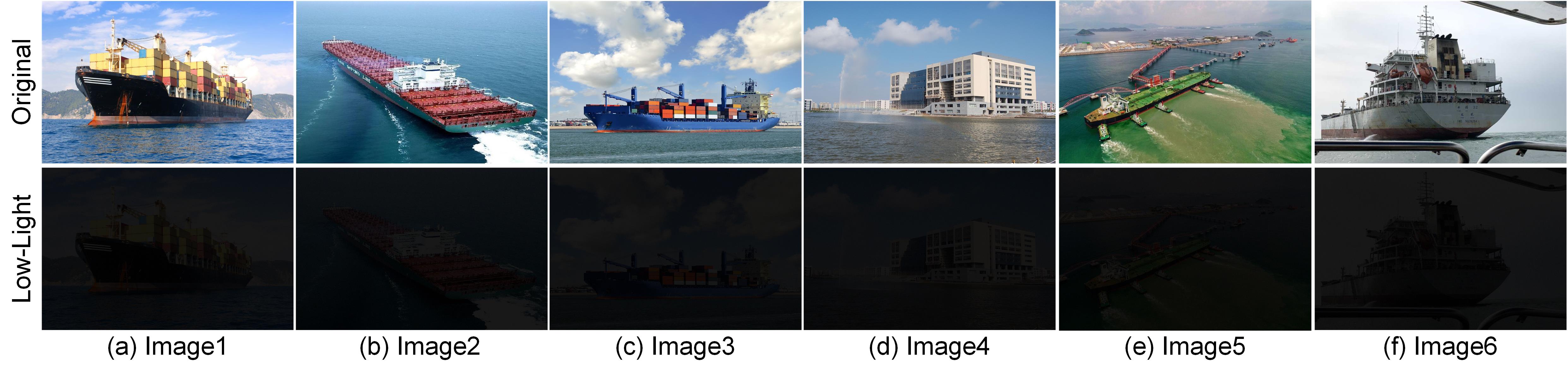}
	\caption{Six synthetic noisy low-light maritime images generated by multiplying the V channel of original sharp images with a coefficient of $\mathcal{C}=0.1$, and then by adding additive Gaussian noise with variance of $\mathcal{V}=25$.}
	\label{Fig4}
\end{figure*}
\subsection{Detail Boosting on Reflection}
\label{sub-sec:DBR}
According to the assumption of Retinex theory, we can easily generate the coarse reflection map $R$ from the optimized illumination map $\hat{L}$ and low-light image $I$ via Eq. (\ref{Eq:Retinex}), i.e.,
\begin{equation*}
R = \frac{I}{\hat{L}} = \frac{I_0 + N}{\hat{L}},
\end{equation*}
where $I_0$ and $N$, respectively, denote the latent noise-free image and unwanted noise. The existence of noise $N$ easily degrades the quality of reflection map leading to unsatisfactory imaging performance. Motivated by previous studies \cite{PTCMDB}, we tend to adopt the guided filter \cite{GuidedF} to effectively boost the details existed in reflection map. In particular, the blurred base layer $B$ is defined as follows
\begin{equation}\label{Eq:Guidedf}
B = G_g(I) \otimes R, 
\end{equation}
where $\otimes$ denotes the convolution operator, $G_g (\cdot)$ is a guided kernel related to the input low-light image $I$ (i.e., guided image) with the local window radius being $15$ and regularization parameter being $10^{-5}$. Please refer to \cite{GuidedF} for more details on guided filter. The details layer $D$ can be defined as follows
\begin{equation}\label{Eq:Detail Boosting1}
D(\mathrm{x}) = R(\mathrm{x}) - B(\mathrm{x}).
\end{equation}

In this work, the enhanced reflection map $\hat{R}$ can be obtained by summing the blurred base layer and weighted details layer, i.e.,
\begin{equation}\label{Eq:Detail Boosting2}
\hat{R}(\mathrm{x}) = B(\mathrm{x}) + \kappa D(\mathrm{x}),
\end{equation}
where $\kappa > 0$ is a weighting parameter. The final enhanced image $\bar{I}$ can be accordingly obtained by multiplying the enhanced reflection map $\hat{R}$ and adjusted illumination map $\hat{L}_G$, i.e.,
\begin{equation}\label{Eq:Detail Boosting3}
\bar{I}(\mathrm{x}) = \hat{R}(\mathrm{x}) \circ \hat{L}_G(\mathrm{x}).
\end{equation}

The proposed detail boosting strategy is capable of preserving the important geometrical structures and suppressing the unwanted artifacts during reflection optimization.    
\subsection{Deep Learning-Based Blind Image Denoising}
\label{sub-sec:DLBBID}

It is obvious that it is computationally difficult to accurately estimate the level fo random noise in practical imaging conditions. Therefore, the current non-blind denoising networks often fail to effectively reduce the unwanted noise easily leading to detail loss or noise residue. To further enhance image quality, we tend to introduce the blind denoising network to blindly remove the unwanted noise existed in enhanced images. To the best of our knowledge, no research has been conducted on blind denoising for low-light image enhancement thus far. The network architecture and loss function for our blind denoising network will be detailedly discussed in this subsection.
\begin{figure*}[t]
	\centering
	\includegraphics[width=1.0\linewidth]{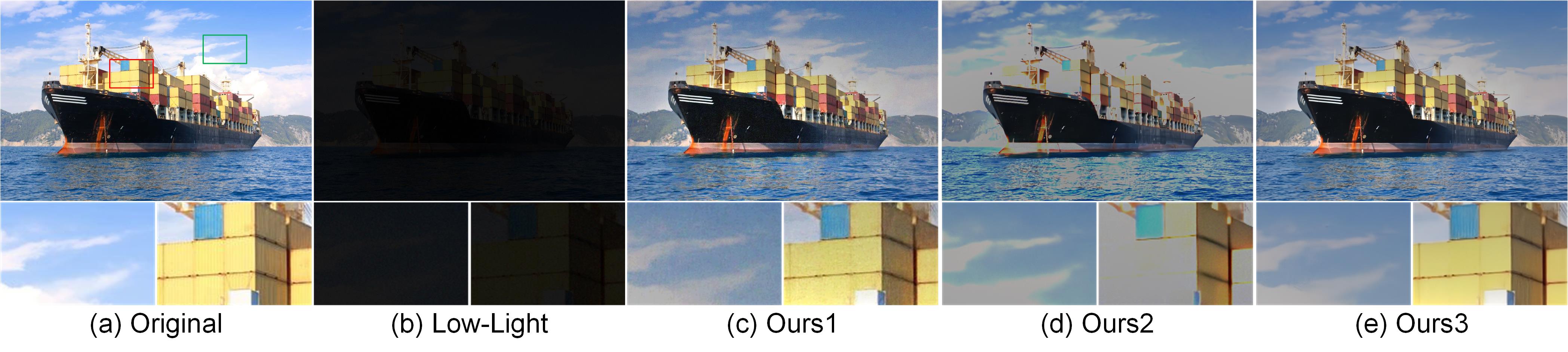}
	\caption{Comparisons of synthetic experiments on Image1 in Fig. \ref{Fig4}. From left to right: (a) original sharp image, (b) low-light image, and enhanced images generated by (c) denoising before image enhancement (i.e., Ours1), (d) denoising the separated reflection map (i.e., Ours2), and (e) denoising after image enhancement (i.e., Ours3), respectively.}
	\label{Fig5}
\end{figure*}
\begin{figure*}[t]
	\centering
	\includegraphics[width=1.0\linewidth]{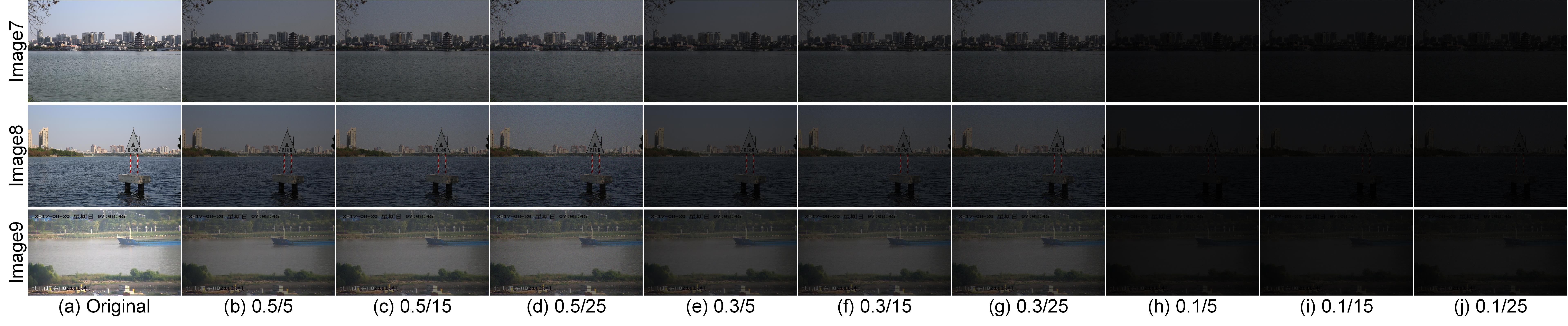}
	\caption{Three different original sharp images (i.e., Image7, Image8 and Image9) adopted to manually generate synthetic low-light images with additive Gaussian noise. From left to right: (a) original images, and noisy low-light images with (b) $\mathcal{C}=0.5$/$\mathcal{V}=5$, (c) $\mathcal{C}=0.5$/$\mathcal{V}=15$, (d) $\mathcal{C}=0.5$/$\mathcal{V}=25$, (e) $\mathcal{C}=0.3$/$\mathcal{V}=5$, (f) $\mathcal{C}=0.3$/$\mathcal{V}=15$, (g) $\mathcal{C}=0.3$/$\mathcal{V}=25$, (h) $\mathcal{C}=0.1$/$\mathcal{V}=5$, (i) $\mathcal{C}=0.1$/$\mathcal{V}=15$, and (j) $\mathcal{C}=0.1$/$\mathcal{V}=25$, respectively.}
	\label{Fig6}
\end{figure*}
\subsubsection{Network Architecture} 
Inspired by previous work \cite{CBDNet}, the blind denoising network introduced in this work is composed of a noise estimation sub-network (E-Net) and a non-blind denoising sub-network (D-Net), shown in Fig. \ref{Blind}. E-Net takes the enhanced image containing noise $\bar{I}$ as input and the estimated noise level map $\hat{\sigma}=\mathcal{F}_E(\bar{I}, N_E)$ as output with $\mathcal{F}_E$ and $N_E$ being the process of estimating the noise level map and E-Net parameters, respectively. D-Net takes $\bar{I}$ and $\hat{\sigma}$ as input and the denoised image $\hat{I}=\mathcal{F}_D (\bar{I},\hat{\sigma}, N_D)$ as output with $ \mathcal{F}_D $ and $N_D$ being the process of estimating the noise-free image and D-Net parameters, respectively. Since $\bar{I}$ and $\hat{\sigma}$ have the same size, E-Net introduces a five-layer fully convolutional network to obtain $\hat{\sigma}$, only containing Convolution (Conv) \cite{Conv} and Rectified Linear Units (ReLU) \cite{ReLU}. In each convolution layer, the number of feature channels is set as $32$, and the size of all filters in E-Net is $3 \times 3$. The ReLU activation function is deployed after each Conv layer. D-Net uses the residual learning \cite{Resiadual L} strategy that first estimates the residual map $\mathcal{R}$ and then obtains $\hat{I} = \bar{I}+\mathcal{R}(\bar{I}, \hat{\sigma}, N_D)$. To improve the network performance, D-Net is generated by modifying the 16-layer U-Net structure \cite{UNet}. In particular, D-Net uses the symmetric skip connections, transposes the convolutions to expand the receiving domain, and uses the multi-scale information. Besides, to further improve the network receptive field, four Conv layers in the middle are revised to dilated convolutions, and their dilated rates are set as $2$, $4$, $8$, and $16$ in order. The size of all filters in D-Net is $3 \times 3$, and the ReLU activation function is deployed after each Conv layer except the last one. 
\subsubsection{Loss Function}
To enhance the network robustness, three loss functions are introduced to constrain the estimated noise level map $\hat{\sigma}$ and the denoised image $\hat{I}$. Recent studies \cite{CBDNet} have shown that non-blind denoising networks were often more sensitive if $\hat{\sigma}(\mathrm{x}) < \tilde{\sigma} (\mathrm{x})$, and more robust if $\hat{\sigma}(\mathrm{x}) > \tilde{\sigma} (\mathrm{x})$ with $ \tilde{\sigma} (\mathrm{x})$ being the ground-truth noise level. To robustly predict the accurate map of noise level, the asymmetric MSE $\mathcal{L}_{asymm}$, total variation $\mathcal{L}_{TV}$, and structural similarity $\mathcal{L}_{SSIM}$ are simultaneously considered as loss function to constrain the estimation of $\hat{\sigma}$. The definition of $\mathcal{L}_{asymm}$ is given by
\begin{equation}\label{Eq:Loss asymm}
\mathcal{L}_{asymm} = \sum_{\mathrm{x} \in \Omega} \left| \alpha - \Bbbk_{(\hat{\sigma} ( \mathrm{x} ) - \tilde{\sigma} ( \mathrm{x} )) < 0} \right| \left( \hat{\sigma} ( \mathrm{x} ) - \tilde{\sigma} ( \mathrm{x} ) \right)^{2},
\end{equation}
where $\Bbbk_{\omega} = 1$ if $\omega < 0$ and $0$ otherwise. As discussed in \cite{CBDNet}, the selection of $\alpha \in (0, 0.5)$ is able to make the network generalize well to realistic noise with more penalty to under-estimation error. The loss function $\mathcal{L}_{TV}$ is defined as follows
\begin{equation}\label{Eq:Loss TV}
\mathcal{L}_{TV} = \sum\limits_{ \mathrm{x} \in \Omega}(\nabla_{h}\hat{\sigma} ( \mathrm{x} ))^{2} + (\nabla_{v}\hat{\sigma} ( \mathrm{x} ))^{2},
\end{equation} 
where $\mathrm{x} \in \Omega$, $\nabla_{h}$ and $\nabla_{v}$ represent the operators of the horizontal and vertical gradients, respectively. To preserve the important geometrical structures in final enhanced images, $\mathcal{L}_{SSIM}$ is also adopted as the loss function, i.e.,
\begin{equation}\label{Eq:loss SSIM}
\mathcal{L}_{SSIM} = \sum_{\mathrm{x} \in \Omega} 1 - \mathrm{SSIM} (\hat{I} ( \mathrm{x} ), \tilde{I} ( \mathrm{x} )).
\end{equation}
where $ \tilde{I} $ is the ground-truth image. In Eq. (\ref{Eq:loss SSIM}), the formulation of SSIM is mathematically defined as follows
\begin{equation}\label{Eq:SSIM}
\mathrm{SSIM} (\hat{I},\tilde{I}) = \frac{(2\mu_{\hat{I}} \mu_{\tilde{I}} + c_1) (2\sigma_{\hat{I}\tilde{I}}+c_2)} {(\mu^{2}_{\hat{I}} + \mu^{2}_{\tilde{I}}+c_1) (\sigma^{2}_{\hat{I}} + \sigma^{2}_{\tilde{I}}+c_2)},
\end{equation}
where $\mu_{\hat{I}}$ and $\mu_{\tilde{I}}$ denote the local mean values, $\sigma_{\hat{I}}$ and $\sigma_{\tilde{I}}$ represent the standard deviations, $\sigma_{\hat{I}\tilde{I}}$ is the covariance value, $c_1$ and $c_2$ are constant parameters. More details on the definition of SSIM can be found in \cite{SSIM}. To sum up, the overall loss function of our blind denoising network can be written as follows
\begin{equation}\label{Eq:Loss whole}
\mathcal{L} = \mathcal{L}_{SSIM} + \lambda_{asymm} \mathcal{L}_{asymm} + \lambda_{TV} \mathcal{L}_{TV},
\end{equation}
where $\lambda_{asymm}$ and $\lambda_{TV}$ represent the trade-off parameters for the asymmetric loss $\mathcal{L}_{asymm}$ and total variation loss $\mathcal{L}_{TV}$, respectively.

During the training of our introduced network, the network parameters of each layer are obtained by minimizing Eq. (\ref{Eq:Loss whole}). We hold the view that the effect of blind denoising network is closely related to the selection of parameters. By manually performing exhaustive numerical experiments, we optimally selected the parameters $\alpha=0.3$, $\lambda_{asymm}=0.5$ and $\lambda_{TV}=0.005$ according to the received enhancement performance.
%
%
%
\section{Experimental Results and Discussion}
\label{sec:ERD}
In this section, we first investigate the influence of deep blind denoising on image enhancement. Furthermore, our method will be compared with several state-of-the-art low-light enhancement methods. Experiments on both synthetic and realistic low-light maritime images will be performed to demonstrate the effectiveness of the proposed method. Finally, we will perform the comparisons of running time for several image enhancement methods under different experimental conditions.
\begin{figure*}[t]
	\centering
	\includegraphics[width=1.0\linewidth]{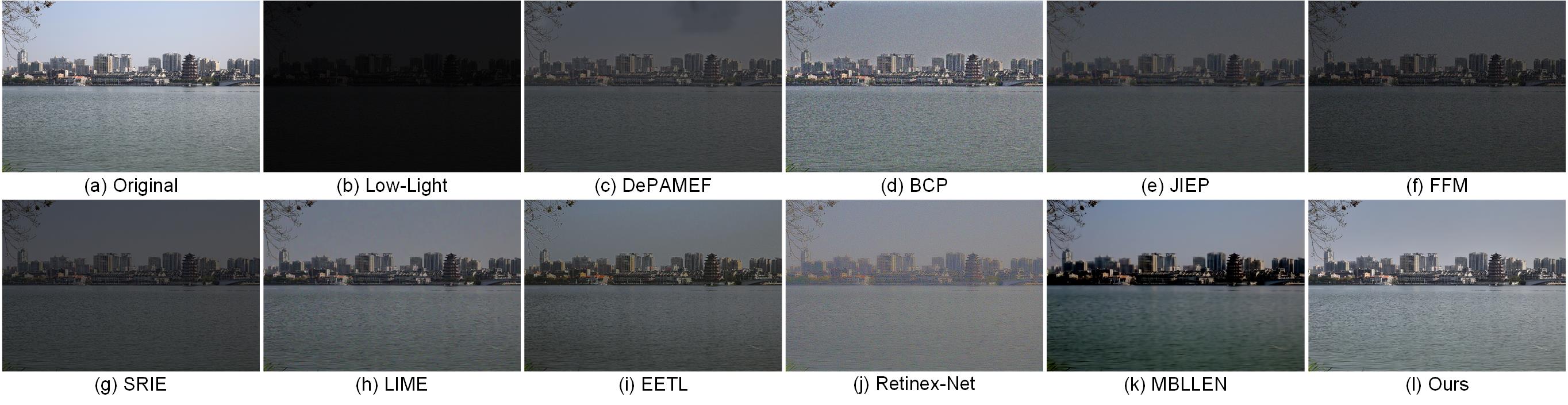}
	\caption{Comparisons of synthetic enhancement experiments on Image7 in Fig. \ref{Fig6}. From top-left to bottom-right: (a) original sharp image, (b) synthetic low-light image with $\mathcal{C}=0.1$/$\mathcal{V}=25$, and enhanced images generated by (c) DePAMEF \cite{DePAMEF}, (d) BCP \cite{BCP}, (e) JIEP \cite{JIEP}, (f) FFM \cite{FFM}, (g) SRIE \cite{SRIE}, (h) LIME \cite{LIME}, (i) EETL \cite{EETL}, (j) Retinex-Net \cite{Retinex-Net}, (k) MBLLEN \cite{MBLLEN}, and (l) Our method, respectively.}
	\label{Fig7}
\end{figure*}
\begin{figure*}[t]
	\centering
	\includegraphics[width=1.0\linewidth]{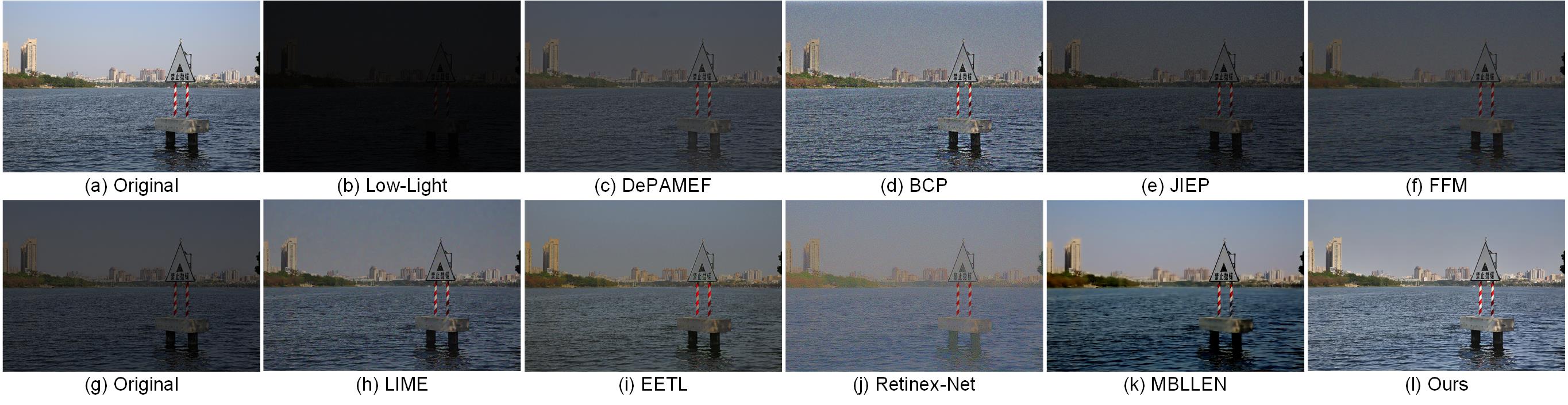}
	\caption{Comparisons of synthetic enhancement experiments on Image8 in Fig. \ref{Fig6}. From top-left to bottom-right: (a) original sharp image, (b) synthetic low-light image with $\mathcal{C}=0.1$/$\mathcal{V}=25$, and enhanced images generated by (c) DePAMEF \cite{DePAMEF}, (d) BCP \cite{BCP}, (e) JIEP \cite{JIEP}, (f) FFM \cite{FFM}, (g) SRIE \cite{SRIE}, (h) LIME \cite{LIME}, (i) EETL \cite{EETL}, (j) Retinex-Net \cite{Retinex-Net}, (k) MBLLEN \cite{MBLLEN}, and (l) Our method, respectively.}
	\label{Fig8}
\end{figure*}
\begin{table*}[t]
	\setlength{\tabcolsep}{5pt}
	\centering
	\caption{PSNR, SSIM, FSIM, and LOE comparisons (mean$\pm$std) of three blind denoising strategies on six different test images visually shown in Fig. \ref{Fig4}.}
	\begin{tabular}{|c|c|c|c|c|}
		\hline
		Methods     & PSNR        & SSIM          & FSIM  		& LOE  \\ \hline \hline
		Low-Light 	& $~~5.3608 \pm 1.8310$ 	&$0.0975 \pm 0.0258$ 	&$0.5884 \pm 0.0549$  &$214 \pm 193$ \\ \hline
		Ours1       & {\color{green}$18.3593 \pm 2.5844$} 	&{\color{green}$0.8556 \pm 0.0620$}	&{\color{green}$0.9435 \pm 0.0204$} &{\color{blue}$147 \pm 131$} \\ \hline
		Ours2       & {\color{blue}$17.2319 \pm 4.2438$}	&{\color{blue}$0.8224 \pm 0.0392$}	&{\color{blue}$0.8812 \pm 0.0303$} &{\color{green}$143 \pm 117$} \\ \hline
		Ours3       & {\color{red}$18.3666 \pm 4.0650$} & {\color{red}$0.9099 \pm 0.0386$} & {\color{red}$0.9611 \pm 0.0063$} &{\color{red}$136 \pm 100$} \\ \hline

	\end{tabular} \label{table1}
\end{table*}
\begin{figure*}[t]
	\centering
	\includegraphics[width=1.0\linewidth]{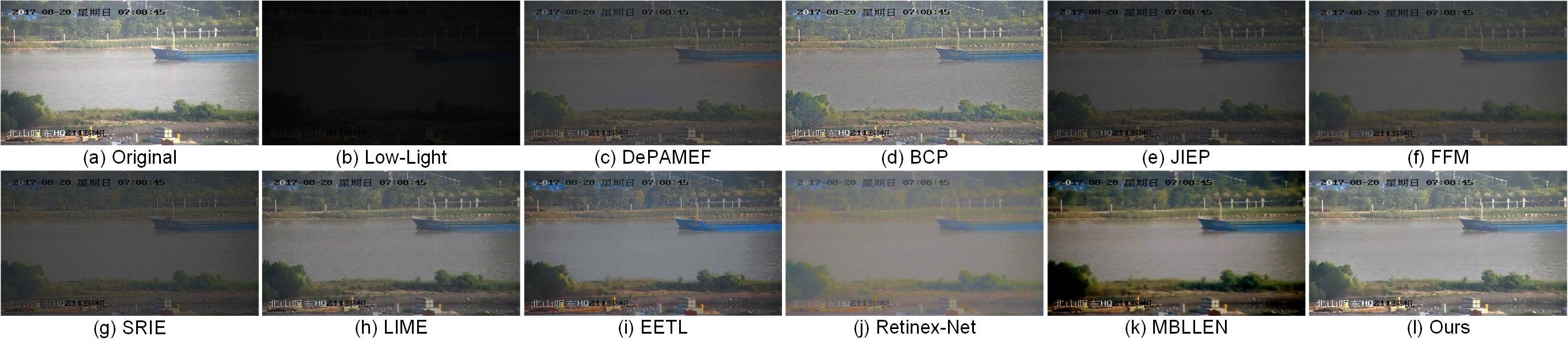}
	\caption{Comparisons of synthetic enhancement experiments on Image9 in Fig. \ref{Fig6}. From top-left to bottom-right: (a) original sharp image, (b) synthetic low-light image with $\mathcal{C}=0.1$/$\mathcal{V}=25$, and enhanced images generated by (c) DePAMEF \cite{DePAMEF}, (d) BCP \cite{BCP}, (e) JIEP \cite{JIEP}, (f) FFM \cite{FFM}, (g) SRIE \cite{SRIE}, (h) LIME \cite{LIME}, (i) EETL \cite{EETL}, (j) Retinex-Net \cite{Retinex-Net}, (k) MBLLEN \cite{MBLLEN}, and (l) Our method, respectively.}
	\label{Fig9}
\end{figure*}
\subsection{Comparisons with Other Enhancement Methods}
\label{sub-sec:COEM}
Our proposed method will be compared with ten state-of-the-art methods including seven traditional methods and three deep learning-based methods.
\begin{itemize}
	\item {DeHZ}: \textit{Dehazing-Based Method} \cite{DeHZ}. This method assumes that the inverted low-light image looks similar to the hazy image. It thus can directly use the DCP-based method to deal with the inverted image (i.e., dehazing). The final enhanced image is accordingly obtained by inverting the dehazed image again.
	\item {DePAMEF}: \textit{Multi-Exposure Image Fusion Dehazing-Based Method} \cite{DePAMEF}. DePAMEF proposes a new single-image dehazing solution based on the adaptive structure decomposition integrated multi-exposure image fusion. In imaging experiments, we first invert the low-light images and then adopt this dehazing method to deal with the inverted images. The final enhanced images are achieved by converting the dehazed images again.
	\item {BCP}: \textit{Bright Channel Prior} \cite{BCP}. To handle the problem of low-light image enhancement, a hybrid regularized variational model is proposed by introducing the bright channel prior, which can eliminate the black halo and suppress the over-enhancement. The resulting minimization problem is effectively solved using an alternating direction optimization method.
	\item {JIEP}: \textit{Joint Intrinsic-Extrinsic Prior} \cite{JIEP}. This method takes full account of the internal characteristics (i.e., shape and texture) and the external environment (i.e., illumination). The proposed joint intrinsic-extrinsic prior model is capable of robustly estimating both illumination and reflection maps.
	\item {FFM}: \textit{Fractional-Order Fusion Model} \cite{FFM}. This model proposes a fractional-order mask and a fusion framework to enhance the low-light images. It can achieve a good trade-off between contrast improvement, detail enhancement, and artifacts suppression.
	\item {SRIE}: \textit{Simultaneous Reflectance and Illumination Estimation} \cite{SRIE}. Based on the analysis of illumination map structure, a new weighted variational model is proposed for better prior representation. This method can not only preserve the estimated reflectance with more details, but also suppress noise to some extent.
	\item {LIME}: \textit{Low-Light Image Enhancement} \cite{LIME}. LIME proposes a structure-aware smoothing model to refine the illumination map which is further adjusted using a gamma correction. By adopting BM3D \cite{BM3D} to suppress the unwanted noise, the final enhanced image can be generated accordingly.
	\item {EETL}: \textit{End-to-End Transformation Learning} \cite{EETL}. It is an end-to-end deep learning method that can convert ordinary photos into DSLR quality images. Furthermore, EETL could adopt the residual convolutional neural network \cite{Resiadual L} and the composite perceptual error function to improve both color rendition and image sharpness.
	\item {Retinex-Net}: \textit{Deep Retinex Decomposition for Low-Light Enhancement} \cite{Retinex-Net}. Retinex-Net combines Retinex theory and deep learning to construct two networks, i.e., Decom-Net for image decomposition and Enhance-Net for illumination adjustment. Furthermore, BM3D \cite{BM3D} is introduced to suppress the unsatisfactory noise existed in reflection map. The enhanced image is finally obtained by multiplying the denoised reflection and adjusted illumination maps.
	\item {MBLLEN}: \textit{Multi-Branch Low-Light Enhancement Network} \cite{MBLLEN}. MBLLEN decomposes the image enhancement problem into sub-problems related to different feature levels, which can be solved respectively to produce the final output via multi-branch fusion.
\end{itemize}

Both synthetic and realistic images will be adopted to evaluate the enhancement performance of these competing methods in terms of quantitative and qualitative evaluations. Four popular full-reference image quality assessment methods, i.e., peak-signal-to-noise ratio (PSNR) \cite{MSE}, structural similarity (SSIM) \cite{SSIM}, feature similarity (FSIM) \cite{FSIM}, and lightness-order-error (LOE) \cite{LOE} are introduced to evaluate the enhancement quality by comparing the enhanced image with the ground-truth version. Meanwhile, two popular no-reference image quality assessment methods, i.e., natural image quality evaluator (NIQE) \cite{NIQE} and blind tone-mapped quality index (BTMQI) \cite{BTMQI} are also employed to perform blind image quality evaluation in realistic experiments. We refer interested readers to Refs. \cite{MSE, SSIM, FSIM, LOE, NIQE, BTMQI} and the references therein for more details on calculations of PSNR, SSIM, FSIM, LOE, NIQE, and BTMQI. Theoretically, higher values of PSNR, SSIM, FSIM, and lower values of LOE, NIQE, BTMQI normally indicate better imaging performance for low-light image enhancement.

\begin{table*}[t]
	\setlength{\tabcolsep}{2.3pt}
	\centering
	\caption{PSNR comparisons (mean$\pm$std) of various image enhancement methods on all test images shown in Fig. \ref{Fig6}.}
	\begin{tabular}{|c|c|c|c|c|c|c|c|c|c|}
		\hline
		\multirow{2}{*}{Methods}    
		& \multicolumn{3}{|c|}{$\mathcal{C}=0.1$}            & \multicolumn{3}{|c|}{$\mathcal{C}=0.3$}         &\multicolumn{3}{|c|}{$\mathcal{C}=0.5$}  \\ \cline{2-10}
		& $\mathcal{V}=5$& $\mathcal{V}=15$& $\mathcal{V}=25$& $\mathcal{V}=5$& $\mathcal{V}=15$& $\mathcal{V}=25$& $\mathcal{V}=5$& $\mathcal{V}=15$& $\mathcal{V}=25$ \\ \hline \hline
		DePAMEF \cite{DePAMEF}  &$9.99  \pm 0.47$ &$9.95  \pm 0.44$ &$9.87  \pm 0.43$ &$16.09 \pm 0.97$ &$15.85 \pm 0.64$ &$14.97 \pm 0.47$ &$19.38 \pm 1.31$ &$17.12 \pm 0.72$ &$14.97 \pm 0.17$\\ \hline
		BCP \cite{BCP}  &${\color{green}19.05 \pm 0.34}$ &${\color{green}18.11 \pm 0.48}$ &${\color{green}16.64 \pm 0.34}$ &${\color{blue}18.32 \pm 0.32}$ &${\color{blue}17.88 \pm 0.32}$ &${\color{blue}16.67 \pm 0.32}$ &$18.30 \pm 0.32$ &$17.89 \pm 0.32$ &$16.68 \pm 0.32$\\ \hline
		JIEP \cite{JIEP}  &$9.17  \pm 0.34$ &$8.88  \pm 0.32$ &$8.61  \pm 0.29$ &$14.34 \pm 0.51$ &$13.45 \pm 0.40$ &$12.66 \pm 0.31$ &${\color{green}20.08 \pm 0.75}$ &$18.04 \pm 0.50$ &$16.26 \pm 0.28$\\ \hline
		FFM \cite{FFM}  &$10.18 \pm 0.39$ &$9.97  \pm 0.38$ &$9.71  \pm 0.36$ &$13.60 \pm 0.51$ &$13.14 \pm 0.46$ &$12.59 \pm 0.40$ &$16.42 \pm 0.58$ &$15.72 \pm 0.49$ &$14.85 \pm 0.40$\\ \hline
		SRIE \cite{SRIE}  &$9.15  \pm 0.35$ &$8.88  \pm 0.33$ &$8.63  \pm 0.30$ &$14.21 \pm 0.51$ &$13.44 \pm 0.43$ &$12.69 \pm 0.33$ &$19.89 \pm 0.70$ &$18.04 \pm 0.51$ &$16.33 \pm 0.32$\\ \hline
		LIME \cite{LIME}  &$14.44 \pm 0.59$ &$13.91 \pm 0.52$ &$13.25 \pm 0.45$ &${\color{red}22.66 \pm 2.34}$ &${\color{green}20.77 \pm 1.37}$ &${\color{green}18.53 \pm 0.80}$ &$19.19 \pm 0.88$ &${\color{blue}18.96 \pm 1.01}$ &${\color{blue}17.69 \pm 0.79}$\\ \hline
		EETL \cite{EETL}  &$13.22 \pm 1.28$ &$12.81 \pm 0.87$ &$13.17 \pm 1.44$ &$16.23 \pm 1.31$ &$16.06 \pm 1.58$ &$15.54 \pm 1.22$ &${\color{blue}19.98 \pm 1.47}$ &${\color{green}19.60 \pm 1.64}$ &${\color{green}18.98 \pm 1.54}$\\ \hline
		Retinex-Net \cite{Retinex-Net}  &${\color{blue}16.47 \pm 0.41}$ &${\color{blue}16.35 \pm 0.38}$ &${\color{blue}16.13 \pm 0.35}$ &$16.07 \pm 0.65$ &$16.04 \pm 0.65$ &$15.94 \pm 0.62$ &$15.05 \pm 0.67$ &$15.19 \pm 0.66$ &$15.28 \pm 0.61$\\ \hline
		MBLLEN \cite{MBLLEN}  &$14.06 \pm 2.20$ &$14.12 \pm 2.28$ &$14.23 \pm 2.35$ &$14.22 \pm 2.65$ &$14.31 \pm 2.66$ &$14.38 \pm 2.62$ &$14.30 \pm 2.77$ &$14.40 \pm 2.79$ &$14.44 \pm 2.71$\\ \hline
		Ours  &${\color{red}21.04 \pm 2.05}$ &${\color{red}21.06 \pm 1.92}$ &${\color{red}20.36 \pm 1.47}$ &${\color{green}22.01 \pm 2.08}$ &${\color{red}21.96 \pm 1.96}$ &${\color{red}21.10 \pm1.51}$ &${\color{red}22.88 \pm 2.11}$ &${\color{red}22.81 \pm 1.99}$ &${\color{red}21.81 \pm 1.48}$\\ \hline
		
	\end{tabular} \label{table2}
\end{table*}

\begin{table*}[t]
	\setlength{\tabcolsep}{0.4pt}
	\centering
	\caption{SSIM comparisons (mean$\pm$std) of various image enhancement methods on all test images shown in Fig. \ref{Fig6}.}
	\begin{tabular}{|c|c|c|c|c|c|c|c|c|c|}
		\hline
		\multirow{2}{*}{Methods}    
		& \multicolumn{3}{|c|}{$\mathcal{C}=0.1$}            & \multicolumn{3}{|c|}{$\mathcal{C}=0.3$}         &\multicolumn{3}{|c|}{$\mathcal{C}=0.5$}  \\ \cline{2-10}
		& $\mathcal{V}=5$& $\mathcal{V}=15$& $\mathcal{V}=25$& $\mathcal{V}=5$& $\mathcal{V}=15$& $\mathcal{V}=25$& $\mathcal{V}=5$& $\mathcal{V}=15$& $\mathcal{V}=25$ \\ \hline \hline
		DePAMEF \cite{DePAMEF}  &$0.622 \pm 0.027$ &$0.566 \pm 0.017$ &$0.497 \pm 0.020$ &$0.865 \pm 0.029$ &$0.694 \pm 0.052$ &$0.560 \pm 0.052$ &$0.798 \pm 0.048$ &$0.594 \pm 0.060$ &$0.472 \pm 0.049$\\ \hline
		BCP \cite{BCP}  &${\color{green}0.886 \pm 0.007}$ &${\color{green}0.748 \pm 0.026}$ &$0.623 \pm 0.038$ &${\color{blue}0.885 \pm 0.014}$ &$0.748 \pm 0.024$ &$0.625 \pm 0.037$ &$0.886 \pm 0.014$ &$0.749 \pm 0.024$ &$0.626 \pm 0.038$\\ \hline
		JIEP \cite{JIEP}  &$0.565 \pm 0.014$ &$0.503 \pm 0.013$ &$0.429 \pm 0.019$ &$0.868 \pm 0.009$ &$0.726 \pm 0.026$ &$0.594 \pm 0.041$ &${\color{green}0.926 \pm 0.003}$ &$0.753 \pm 0.039$ &$0.613 \pm 0.052$\\ \hline
		FFM \cite{FFM}  &$0.558 \pm 0.032$ &$0.503 \pm 0.029$ &$0.435 \pm 0.031$ &$0.796 \pm 0.022$ &$0.684 \pm 0.023$ &$0.569 \pm 0.037$ &$0.882 \pm 0.009$ &$0.737 \pm 0.032$ &$0.602 \pm 0.047$\\ \hline
		SRIE \cite{SRIE}  &$0.559 \pm 0.031$ &$0.496 \pm 0.014$ &$0.423 \pm 0.009$ &$0.848 \pm 0.024$ &$0.714 \pm 0.013$ &$0.587 \pm 0.032$ &$0.918 \pm 0.011$ &$0.750 \pm 0.031$ &$0.610 \pm 0.046$\\ \hline
		LIME \cite{LIME}  &${\color{blue}0.840 \pm 0.030}$ &$0.723 \pm 0.043$ &$0.611 \pm 0.050$ &${\color{green}0.930 \pm 0.018}$ &$0.766 \pm 0.050$ &$0.624 \pm 0.061$ &$0.905 \pm 0.028$ &$0.723 \pm 0.062$ &$0.576 \pm 0.066$\\ \hline
		EETL \cite{EETL}  &$0.774 \pm 0.041$ &$0.724 \pm 0.028$ &${\color{blue}0.658 \pm 0.046}$ &$0.885 \pm 0.025$ &${\color{green}0.818 \pm 0.028}$ &${\color{blue}0.702 \pm 0.033}$ &${\color{blue}0.925 \pm 0.016}$ &${\color{green}0.835 \pm 0.016}$ &${\color{green}0.708 \pm 0.029}$\\ \hline
		Retinex-Net \cite{Retinex-Net}  &$0.732 \pm 0.061$ &$0.679 \pm 0.046$ &$0.607 \pm 0.030$ &$0.723 \pm 0.136$ &$0.630 \pm 0.100$ &$0.534 \pm 0.064$ &$0.697 \pm 0.133$ &$0.603 \pm 0.088$ &$0.510 \pm 0.052$\\ \hline
		MBLLEN \cite{MBLLEN}  &$0.734 \pm 0.079$ &${\color{blue}0.725 \pm 0.077}$ &${\color{green}0.713 \pm 0.069}$ &$0.797 \pm 0.077$ &${\color{blue}0.786 \pm 0.070}$ &${\color{green}0.727 \pm 0.057}$ &$0.814 \pm 0.077$ &${\color{blue}0.792 \pm 0.069}$ &${\color{blue}0.689 \pm 0.050}$\\ \hline
		Ours  &${\color{red}0.951 \pm 0.013}$ &${\color{red}0.939 \pm 0.005}$ &${\color{red}0.919 \pm 0.006}$ &${\color{red}0.964 \pm 0.011}$ &${\color{red}0.948 \pm 0.002}$ &${\color{red}0.924 \pm 0.009}$ &${\color{red}0.971 \pm 0.009}$ &${\color{red}0.953 \pm 0.001}$ &${\color{red}0.928 \pm 0.011}$\\ \hline	
	\end{tabular} \label{table3}
\end{table*}

\begin{table*}[t]
	\setlength{\tabcolsep}{0.4pt}
	\centering
	\caption{FSIM comparisons (mean$\pm$std) of various image enhancement methods on all test images shown in Fig. \ref{Fig6}.}
	\begin{tabular}{|c|c|c|c|c|c|c|c|c|c|}
		\hline
		\multirow{2}{*}{Methods}    
		& \multicolumn{3}{|c|}{$\mathcal{C}=0.1$}            & \multicolumn{3}{|c|}{$\mathcal{C}=0.3$}         &\multicolumn{3}{|c|}{$\mathcal{C}=0.5$}  \\ \cline{2-10}
		& $\mathcal{V}=5$& $\mathcal{V}=15$& $\mathcal{V}=25$& $\mathcal{V}=5$& $\mathcal{V}=15$& $\mathcal{V}=25$& $\mathcal{V}=5$& $\mathcal{V}=15$& $\mathcal{V}=25$ \\ \hline \hline
		DePAMEF \cite{DePAMEF}  &$0.806 \pm 0.005$ &$0.805 \pm 0.005$ &$0.801 \pm 0.004$ &${\color{blue}0.964 \pm 0.007}$ &$0.946 \pm 0.011$ &$0.911 \pm 0.018$ &$0.925 \pm 0.017$ &$0.889 \pm 0.019$ &$0.849 \pm 0.021$\\ \hline
		BCP \cite{BCP}  &${\color{green}0.947 \pm 0.008}$ &${\color{blue}0.935 \pm 0.001}$ &$0.913 \pm 0.006$ &$0.942 \pm 0.011$ &$0.932 \pm 0.002$ &$0.911 \pm 0.004$ &$0.943 \pm 0.011$ &$0.933 \pm 0.003$ &$0.911 \pm 0.005$\\ \hline
		JIEP \cite{JIEP}  &$0.812 \pm 0.001$ &$0.806 \pm 0.006$ &$0.795 \pm 0.008$ &$0.947 \pm 0.013$ &$0.930 \pm 0.007$ &$0.905 \pm 0.007$ &${\color{green}0.973 \pm 0.012}$ &$0.953 \pm 0.006$ &${\color{blue}0.924 \pm 0.012}$\\ \hline
		FFM \cite{FFM}  &$0.745 \pm 0.009$ &$0.750 \pm 0.008$ &$0.752 \pm 0.008$ &$0.887 \pm 0.010$ &$0.883 \pm 0.008$ &$0.873 \pm 0.007$ &$0.944 \pm 0.007$ &$0.933 \pm 0.005$ &$0.912 \pm 0.008$\\ \hline
		SRIE \cite{SRIE}  &$0.794 \pm 0.022$ &$0.790 \pm 0.014$ &$0.782 \pm 0.006$ &$0.928 \pm 0.023$ &$0.916 \pm 0.014$ &$0.895 \pm 0.008$ &$0.966 \pm 0.017$ &$0.949 \pm 0.008$ &$0.921 \pm 0.008$\\ \hline
		LIME \cite{LIME}  &${\color{blue}0.944 \pm 0.015}$ &${\color{green}0.936 \pm 0.013}$ &${\color{green}0.925 \pm 0.011}$ &${\color{red}0.982 \pm 0.008}$ &${\color{green}0.972 \pm 0.009}$ &${\color{green}0.949 \pm 0.012}$ &${\color{blue}0.969 \pm 0.017}$ &${\color{blue}0.955 \pm 0.018}$ &$0.922 \pm 0.019$\\ \hline
		EETL \cite{EETL}  &$0.901 \pm 0.009$ &$0.895 \pm 0.011$ &$0.890 \pm 0.018$ &$0.945 \pm 0.004$ &$0.935 \pm 0.005$ &$0.913 \pm 0.008$ &$0.958 \pm 0.002$ &$0.946 \pm 0.004$ &$0.919 \pm 0.009$\\ \hline
		Retinex-Net \cite{Retinex-Net}  &$0.780 \pm 0.025$ &$0.781 \pm 0.024$ &$0.779 \pm 0.021$ &$0.799 \pm 0.068$ &$0.795 \pm 0.060$ &$0.787 \pm 0.049$ &$0.788 \pm 0.053$ &$0.783 \pm 0.048$ &$0.772 \pm 0.044$\\ \hline
		MBLLEN \cite{MBLLEN}  &$0.910 \pm 0.023$ &$0.912 \pm 0.021$ &${\color{blue}0.915 \pm 0.018}$ &$0.952 \pm 0.008$ &${\color{blue}0.951 \pm 0.006}$ &${\color{blue}0.937 \pm 0.005}$ &$0.961 \pm 0.005$ &${\color{green}0.956 \pm 0.003}$ &${\color{green}0.932 \pm 0.006}$\\ \hline
		Ours  &${\color{red}0.969 \pm 0.018}$ &${\color{red}0.967 \pm 0.016}$ &${\color{red}0.959 \pm 0.012}$ &${\color{green}0.980 \pm 0.012}$ &${\color{red}0.974 \pm 0.011}$ &${\color{red}0.964 \pm 0.009}$ &${\color{red}0.984 \pm 0.009}$ &${\color{red}0.977 \pm 0.008}$ &${\color{red}0.965 \pm 0.007}$\\ \hline
	\end{tabular} \label{table4}
\end{table*}

\begin{table*}[t]
	\setlength{\tabcolsep}{5.9pt}
	\centering
	\caption{LOE comparisons (mean$\pm$std) of various image enhancement methods on all test images shown in Fig. \ref{Fig6}.}
	\begin{tabular}{|c|c|c|c|c|c|c|c|c|c|}
		\hline
		\multirow{2}{*}{Methods}    
		& \multicolumn{3}{|c|}{$\mathcal{C}=0.1$}            & \multicolumn{3}{|c|}{$\mathcal{C}=0.3$}         &\multicolumn{3}{|c|}{$\mathcal{C}=0.5$}  \\ \cline{2-10}
		& $\mathcal{V}=5$& $\mathcal{V}=15$& $\mathcal{V}=25$& $\mathcal{V}=5$& $\mathcal{V}=15$& $\mathcal{V}=25$& $\mathcal{V}=5$& $\mathcal{V}=15$& $\mathcal{V}=25$ \\ \hline \hline
		DePAMEF \cite{DePAMEF}  &$248  \pm 208$ &$249 \pm 207$ &$250 \pm 207$ &${\color{green}143  \pm 55 }$ &${\color{green}168 \pm 122}$ &$244 \pm 209$ &$188 \pm 135$ &$245 \pm 208$ &$247 \pm 208$\\ \hline
		BCP \cite{BCP}  &${\color{blue}228  \pm 194}$ &${\color{blue}242 \pm 211}$ &$245 \pm 209$ &$245  \pm 209$ &$247 \pm 209$ &$247 \pm 208$ &$245 \pm 209$ &$247 \pm 209$ &$247 \pm 208$\\ \hline
		JIEP \cite{JIEP}  &$248  \pm 208$ &$248 \pm 208$ &$248 \pm 208$ &$807  \pm 49 $ &$384 \pm 58 $ &${\color{blue}177 \pm 110}$ &$154 \pm 114$ &$231 \pm 201$ &$243 \pm 210$\\ \hline
		FFM \cite{FFM}  &$248  \pm 208$ &$248 \pm 208$ &$248 \pm 208$ &$1003 \pm 207$ &$642 \pm 251$ &$236 \pm 182$ &$201 \pm 146$ &${\color{blue}170 \pm 130}$ &${\color{blue}238 \pm 211}$\\ \hline
		SRIE \cite{SRIE}  &$248  \pm 208$ &$248 \pm 208$ &$248 \pm 208$ &$1013 \pm 110$ &$845 \pm 48 $ &$333 \pm 57 $ &${\color{blue}152 \pm 87 }$ &$203 \pm 179$ &$239 \pm 212$\\ \hline
		LIME \cite{LIME}  &$852  \pm 208$ &$437 \pm 231$ &${\color{green}199 \pm 108}$ &${\color{red}137  \pm 86 }$ &${\color{blue}209 \pm 175}$ &$243 \pm 209$ &$204 \pm 158$ &$244 \pm 210$ &$247 \pm 209$\\ \hline
		EETL \cite{EETL}  &$904  \pm 195$ &$992 \pm 149$ &$672 \pm 188$ &$376  \pm 191$ &$243 \pm 103$ &${\color{green}157 \pm 103}$ &${\color{red}125 \pm 38 }$ &${\color{green}152 \pm 103}$ &${\color{green}220 \pm 185}$\\ \hline
		Retinex-Net \cite{Retinex-Net}  &${\color{green}180  \pm 67 }$ &${\color{green}132 \pm 78 }$ &${\color{blue}211 \pm 170}$ &$247  \pm 208$ &$248 \pm 208$ &$248 \pm 208$ &$248 \pm 208$ &$248 \pm 208$ &$248 \pm 208$\\ \hline
		MBLLEN \cite{MBLLEN}  &$1023 \pm 183$ &$998 \pm 200$ &$946 \pm 241$ &$913  \pm 312$ &$854 \pm 373$ &$716 \pm 436$ &$864 \pm 366$ &$803 \pm 423$ &$603 \pm 389$\\ \hline
		Ours  &${\color{red}153  \pm 115}$ &${\color{red}131 \pm 83 }$ &${\color{red}120 \pm 61 }$ &${\color{blue}146  \pm 104}$ &${\color{red}116 \pm 62 }$ &${\color{red}115 \pm 53 }$ &${\color{green}142 \pm 96 }$ &${\color{red}109 \pm 56 }$ &${\color{red}99 \pm 43 }$\\ \hline
	\end{tabular} \label{table5}
\end{table*}

\begin{figure*}[t]
	\centering
	\includegraphics[width=1.0\linewidth]{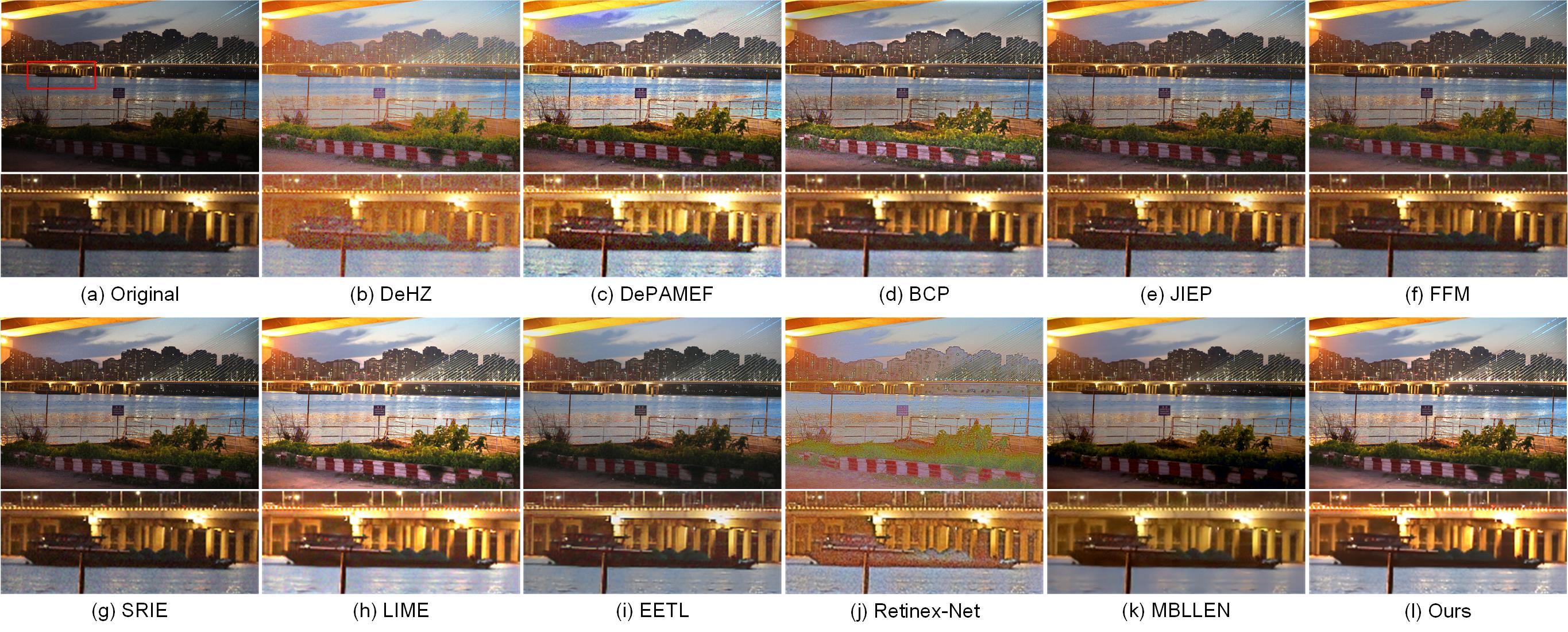}
	\caption{Comparisons of realistic enhancement experiments on Image10. From top-left to bottom-right (NIQE and BTMQI for text in brackets): (a) original low-light image (6.2918, 4.1801), enhanced images yielded by (b) DeHZ \cite{DeHZ} (8.3304, 3.3604), (c) DePAMEF \cite{DePAMEF} (7.0134, 4.3556), (d) BCP \cite{BCP} (4.4693, 3.1606), (e) JIEP \cite{JIEP} (6.6720, 2.7411), (f) FFM \cite{FFM} (6.3025, 2.7312), (g) SRIE \cite{SRIE} (6.1703, 2.5353), (h) LIME \cite{LIME} (4.1544, 4.0813), (i) EETL \cite{EETL} (4.2124, 2.6075), (j) Retinex-Net \cite{Retinex-Net} (5.6486, 3.5642), (k) MBLLEN \cite{MBLLEN} (3.7250, 2.6775), and (l) Our method (3.8786, 2.9748), respectively.}
	\label{Fig10}
\end{figure*}
\begin{figure*}[t]
	\centering
	\includegraphics[width=1.0\linewidth]{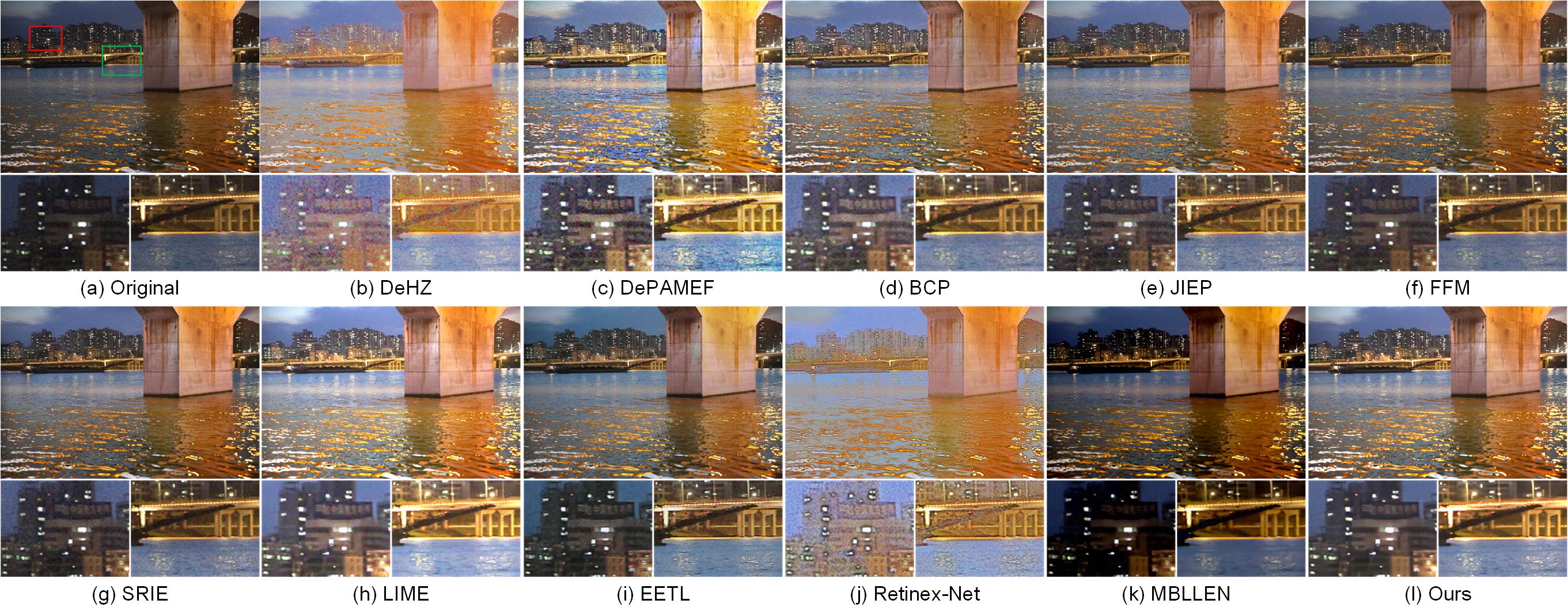}
	\caption{Comparisons of realistic enhancement experiments on Image11. From top-left to bottom-right (NIQE and BTMQI for text in brackets): (a) original low-light image (7.1883, 3.6943), enhanced images yielded by (b) DeHZ \cite{DeHZ} (9.2618, 3.7155), (c) DePAMEF \cite{DePAMEF} (7.6561, 5.1725), (d) BCP \cite{BCP} (5.0638, 2.3075), (e) JIEP \cite{JIEP} (7.4099, 1.9083), (f) FFM \cite{FFM} (7.1266, 1.3251), (g) SRIE \cite{SRIE} (6.6622, 1.9807), (h) LIME \cite{LIME} (4.4044, 4.4075), (i) EETL \cite{EETL} (3.7407, 1.7183), (j) Retinex-Net \cite{Retinex-Net} (5.5713, 3.5419), (k) MBLLEN \cite{MBLLEN} (4.0765, 3.0483), and (l) Our method (3.6826, 2.6244), respectively.}
	\label{Fig11}
\end{figure*}
\begin{figure*}[!]
	\centering
	\includegraphics[width=1.0\linewidth]{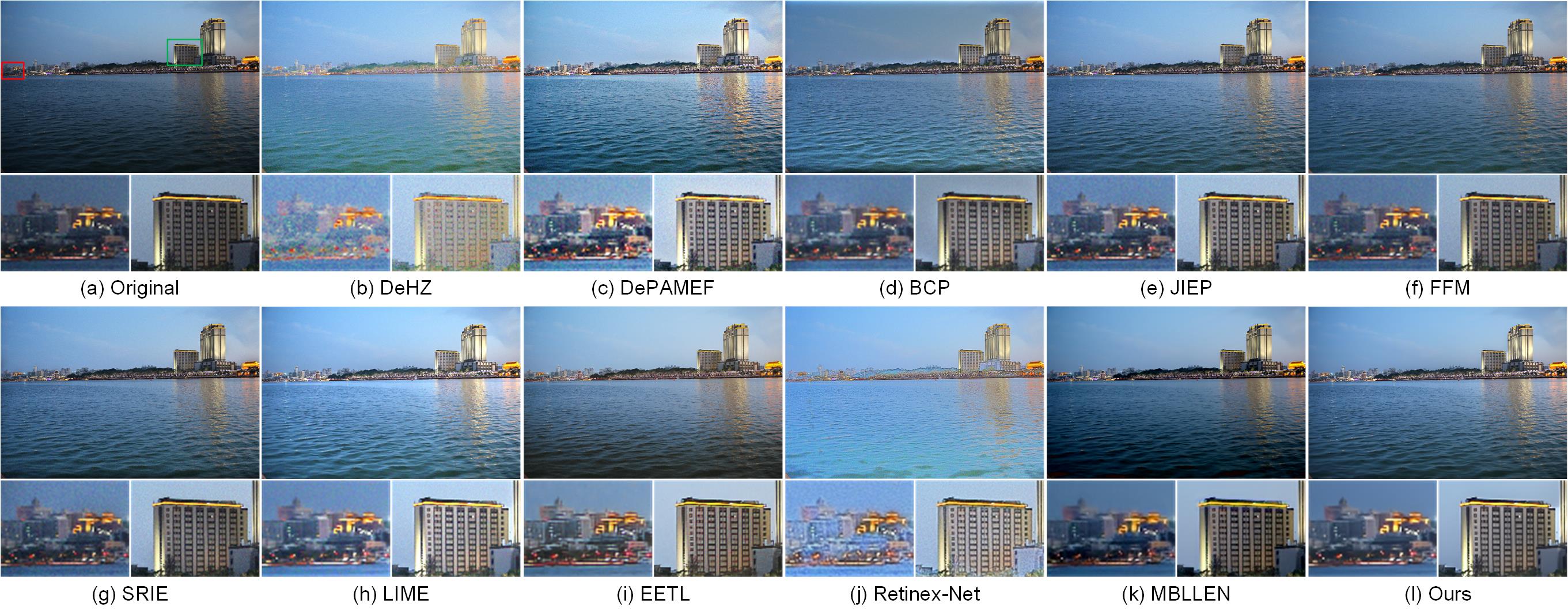}
	\caption{Comparisons of realistic enhancement experiments on Image12. From top-left to bottom-right (NIQE and BTMQI for text in brackets): (a) original low-light image (7.1937, 4.1920), enhanced images yielded by (b) DeHZ \cite{DeHZ} (7.6638, 4.3166), (c) DePAMEF \cite{DePAMEF} (8.0244, 3.0026), (d) BCP \cite{BCP} (5.5081, 2.1617), (e) JIEP \cite{JIEP} (7.5620, 1.1777), (f) FFM \cite{FFM} (7.2641, 1.2374), (g) SRIE \cite{SRIE} (7.2899, 1.4066), (h) LIME \cite{LIME} (5.3992, 3.2538), (i) EETL \cite{EETL} (4.9691, 2.1234), (j) Retinex-Net \cite{Retinex-Net} (7.2269, 4.4598), (k) MBLLEN \cite{MBLLEN} (4.7048, 4.1862), and (l) Our method (5.1281, 1.7688), respectively.}
	\label{Fig12}
\end{figure*}

\subsection{Experimental Settings}
\label{sub-sec:ES}
To guarantee high-quality enhancement results, the input parameters should be selected properly in our experiments, e.g., $\lambda_1$, $\lambda_2$, $\beta_1$, $\beta_2$ and $t$ for illumination optimization, $\gamma_0$ for illumination adjustment and $\kappa$ for detail boosting on reflection. In particular, the regularization parameters $\lambda_1$ and $\lambda_2$ control the trade-offs between the data-fidelity and regularization terms. The penalty parameters $\beta_1$ and $\beta_2$ are of importance in guaranteeing stable solutions. In this work, we propose to adopt the manual method, which experientially tries several values within a predefined range of parameters, to select the proper parameters. To explain how to select these parameters, we performed exhaustive numerical experiments to manually determine the satisfactory selections. According to both quantitative and qualitative evaluations, we manually selected the optimal parameters in our numerical experiments, i.e., $\lambda_1 = 3$, $\lambda_2 = 1$, $\beta_1 = 1$, $\beta_2 = 1$, $t = 0.5$, $\gamma_0 = 1.429$ and $\kappa = 1.3$. These selected parameters for low-light image enhancement were used throughout the rest of this paper. Numerical experiments have shown that the image enhancement results under the current parameter settings were consistently promising. For the sake of fair comparison, other competing enhancement methods were performed by the authors' codes with the optimized parameters.

To improve the imaging performance of blind denoising network, we tend to select $2000$ noise-free images as the dataset. In particular, the synthetic versions are obtained by adding white Gaussian noise with variance $\mathcal{C}$ ranging between $(0, 50)$ on the noise-free images. In our numerical experiments, the learning network is trained for $80$ epochs. To increase the convergence rate, the learning rate of the first $40$ epochs is set to $10^{-3}$ and the learning rate of the last $40$ epochs is set to $10^{-4}$. In each epoch, the dataset is randomly cropped into $34000$ image patches of size $256 \times 256$. All numerical experiments and training network models are conducted in Python $3.7$ and Matlab2019a environment running on a PC with Intel(R) Core (TM) i7-9750H CPU \textcircled{a}2.60GHz and a Nvidia GeForce GTX $2080$ GPU. It takes about $40$ hours to train the blind denoising network with the Pytorch package \cite{Pytorch}. 
%
\subsection{Influence of Deep Blind Denoising on Image Enhancement}
\label{sub-sec:IDDI}
This subsection mainly discusses the influence of deep blind denoising proposed in Section \ref{sub-sec:DLBBID} on final enhancement performance. It is well known the unwanted noise in low-light images could easily be amplified during image enhancement. To overcome this problem, deep blind denoising is able to effectively remove the unsatisfactory artifacts. The important problem in this work is how to adopt the blind denoising network during low-light image enhancement. There are mainly three strategies to incorporate deep blind denoising into our low-light image enhancement framework. For example, the first strategy (a.k.a., Ours1) adopts the deep blind denoising to directly denoise the original image $I$. The denoised low-light image will be enhanced using the refined illumination map (in Section \ref{sub-sec:IA}) and the optimized reflection map (in Section \ref{sub-sec:DBR}). According to the Retinex theory, it is commonly assumed that the estimated reflection map contains random noise. Therefore, the second strategy (a.k.a., Ours2) will adopt the deep blind denoising to handle the reflection map. The refined illumination and denoised reflection maps could then be accordingly combined to generate the final enhanced images. The last strategy (a.k.a., Ours3), shown in Section \ref{sec:PLLIEF}, will use the deep blind denoising to directly optimize the enhanced images to further promote imaging performance.

To investigate the influence of deep blind denoising on image enhancement, six different high-quality maritime images and their noisy low-light versions are illustrated in Fig. \ref{Fig4}. In this work, we first add the white Gaussian noise with variance $\mathcal{V} = 25$ to the original high-quality maritime images. In the second step, we transform the noisy images from RGB color space into HSV color space. The V channel of each image is multiplied by a darkening coefficient $\mathcal{C}$ of 0.1. The noisy low-light versions are synthetically generated by transforming from HSV color space into RGB color space. To objectively evaluate the imaging performance, four quality measures (i.e., PSNR, SSIM, FSIM, and LOE) will be used simultaneously in our numerical experiments.

For the sake of better visual comparison, the synthetic experiments for Ours1, Ours2, and Ours3 on one image are visually shown in Fig. \ref{Fig5}. It can be observed that the enhanced image yielded by Ours1 still suffers from the unwanted noise, leading to visual quality degradation. The reason behind this phenomenon is that the luminance statistics between normal and low-light images are significantly different in essence. It is thus difficult to effectively remove the random noise in low-light regions since the structural information is often ignored. The enhanced image obtained by Ours2 not only has the problem of over enhancement, but also suffers from the serious color distortion. In contrast, our proposed Ours3 can effectively suppress the noise and produce satisfactory visual appearance. The advantage of Ours3 is further confirmed by the values of PSNR, SSIM, FSIM, and LOE summarized in Table \ref{table1}. It is obvious that Ours3 generates the best objective evaluation under all imaging conditions.
\subsection{Experimental Results on Synthetic Maritime Images}
\label{sub-sec:ERSI}
This subsection is devoted to compare our proposed method (i.e., Ours3) with nine popular low-light image enhancement methods, i.e., DePAMEF \cite{DePAMEF}, BCP \cite{BCP}, JIEP \cite{JIEP}, FFM \cite{FFM}, SRIE \cite{SRIE}, LIME \cite{LIME}, EETL \cite{EETL}, Retinex-Net \cite{Retinex-Net}, and MBLLEN \cite{MBLLEN}. Due to the unsatisfactory results, the imaging performance of DeHZ \cite{DeHZ} is not considered in this subsection. The synthetic experiments are performed using three different original sharp images shown in Fig. \ref{Fig6}. To evaluate the stability of our enhancement method, we propose to add the white Gaussian noise with variance $\mathcal{V} \in \{5, 15, 25 \}$ and multiply the darkening coefficient $\mathcal{C} \in \{0.1, 0.3, 0.5 \}$ to synthetically generate the degraded images. To quantitatively evaluate the enhancement performance, four metrics (i.e., PSNR, SSIM, FSIM, and LOE) are adopted simultaneously in our synthetic experiments.

For the sake of better visual comparisons, we only display the image enhancement results under the worst imaging condition, i.e., $\mathcal{V} = 25$ and $\mathcal{C} = 0.1$. The low-light image enhancement results can be visually found in Figs. \ref{Fig7}-\ref{Fig9}. It can be observed that Retinex-Net \cite{Retinex-Net} leads to obvious color distortions and blocking artifacts, resulting in visual quality degradation. The essential reason is that the training datasets adopted may not contain the similar features existed in maritime images to be enhanced. The restored results produced by JIEP \cite{JIEP} and FFM \cite{FFM} obviously suffer from the problem of insufficient enhancement, which causes the loss of fine visual details. In addition, BCP \cite{BCP} can effectively enhance the illumination of low-light images, but it is intractable to effectively suppress the unwanted noise. Due to the structure-aware smoothing model introduced, LIME \cite{LIME} is able to generate higher-quality enhanced images. The residual noise, however, fails to be effectively suppressed in the sky regions, leading to unnatural visual appearance. In contrast, our proposed method is able to enhance the low-light images and suppress the unsatisfactory artifacts in enhanced versions. Our superior performance can be further confirmed by the quantitative results PSNR, SSIM, FSIM, and LOE shown in Table \ref{table2}-\ref{table5}. It can be found that our method outperforms other competing methods under consideration in most of the cases.

\begin{figure*}[t]
	\centering
	\includegraphics[width=1.0\linewidth]{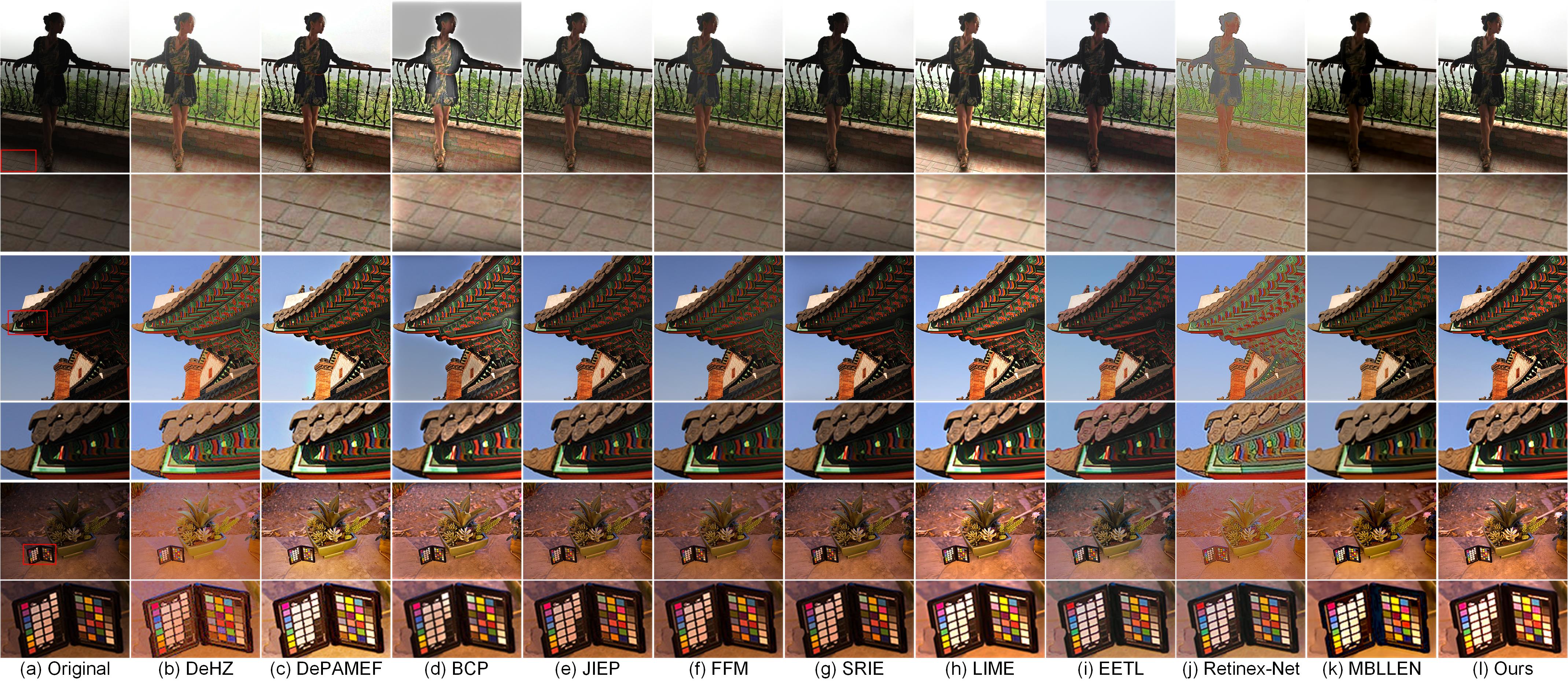}
	\caption{Comparisons of realistic enhancement experiments on Image 13-15 (From top to bottom: Image13, Image14, and Image15). From left to right: (a) original sharp image, enhanced images yielded by (b) DeHZ \cite{DeHZ}, (c) DePAMEF \cite{DePAMEF}, (d) BCP \cite{BCP}, (e) JIEP \cite{JIEP}, (f) FFM \cite{FFM}, (g) SRIE \cite{SRIE}, (h) LIME \cite{LIME}, (i) EETL \cite{EETL}, (j) Retinex-Net \cite{Retinex-Net}, (k) MBLLEN \cite{MBLLEN}, and (l) Our method, respectively.}
	\label{Fig13}
\end{figure*}

\begin{table*}[t]
	\setlength{\tabcolsep}{10.3pt}
	\centering
	\caption{NIQE and BTMQI comparisons of various image enhancement methods on all test images shown in Fig. \ref{Fig13}.}
	\begin{tabular}{|c|c|c|c|c|c|c|c|c|c|}
		\hline
		\multirow{2}{*}{Methods}  
		&\multicolumn{2}{|c|}{Image13} & \multicolumn{2}{|c|}{Image14}  & \multicolumn{2}{|c|}{Image15}& \multicolumn{2}{|c|}{Average} \\ \cline{2-9}
		&NIQE&BMTQI&NIQE&BMTQI&NIQE&BMTQI&NIQE&BMTQI 	 \\ \hline \hline
		Original &$3.0897$ &$5.7702$ &${\color{red}2.9951}$ &$5.7828$ &${\color{green}2.8033}$ &$6.2178$&$2.9627 \pm 0.2064$&$5.9236 \pm 0.3605$ \\ \hline
		DeHZ \cite{DeHZ} &$2.6380$ &$4.1833$ &$3.7053$ &$2.8234$ &$2.4209$ &$4.2360$&$2.9214 \pm 0.9722$&$3.7475 \pm 1.1325$ \\ \hline
		DePAMEF \cite{DePAMEF} &${\color{red}2.3133}$ &$3.0380$ &$3.3417$ &${\color{red}2.6713}$ &$2.8677$ &$3.4564$&$2.8409 \pm 0.7279$&${\color{blue}3.0552 \pm 0.5556}$ \\ \hline
		BCP \cite{BCP} &${\color{green}2.3391}$ &${\color{red}3.2384}$ &$2.9639$ &${\color{blue}2.7706}$ &$3.1484$ &$3.1076$&${\color{blue}2.8171 \pm 0.5998}$&${\color{green}3.0389 \pm 0.3413}$ \\ \hline
		JIEP \cite{JIEP} &$2.5069$ &${\color{green}3.3975}$ &${\color{green}3.0514}$ &$3.5271$ &$2.8540$ &$3.0256$&${\color{green}2.8041 \pm 0.3898}$&$3.3168 \pm 0.3682$ \\ \hline
		FFM \cite{FFM} &$2.6981$ &$4.1116$ &$3.4257$ &$3.1131$ &$2.8465$ &$3.8863$&$2.9901 \pm 0.5437$&$3.7037 \pm 0.7406$ \\ \hline
		SRIE \cite{SRIE} &$2.7101$ &$4.3487$ &$2.8390$ &$4.3869$ &$3.0055$ &$3.6045$&$2.8515 \pm 0.2095$&$4.1134 \pm 0.6238$ \\ \hline
		LIME \cite{LIME} &$3.6220$ &$5.1341$ &$3.3992$ &$2.6305$ &$3.6698$ &${\color{green}2.5349}$&$3.5637 \pm 0.2042$&$3.4332 \pm 2.0843$ \\ \hline
		EETL \cite{EETL} &$4.3326$ &$5.1045$ &$4.4980$ &$4.4911$ &$3.4034$ &${\color{red}2.5015}$&$4.0780 \pm 0.8345$&$4.0323 \pm 1.9244$ \\ \hline
		Retinex-Net \cite{Retinex-Net} &$3.1359$ &$5.4856$ &$4.2532$ &$2.9979$ &${\color{red}2.5126}$ &$3.4347$&$3.3006 \pm 1.2472$&$3.9727 \pm 1.8784$ \\ \hline
		MBLLEN \cite{MBLLEN} &$3.7493$ &$4.9593$ &$3.5385$ &$3.8280$ &$3.6265$ &$3.0036$&$3.6381 \pm 0.1498$&$3.9303 \pm 1.3886$ \\ \hline
		Ours &${\color{blue}2.3640}$ &${\color{blue}3.7057}$ &${\color{blue}3.1856}$ &${\color{green}2.7305}$ &${\color{blue}2.8127}$ &${\color{blue}2.6561}$&${\color{red}2.7874 \pm 0.5818}$&${\color{red}3.0308 \pm 0.8283}$ \\ \hline
	\end{tabular} \label{table6}
\end{table*}
\begin{table}[t]
	\setlength{\tabcolsep}{9.5pt}
	\centering
	\caption{Comparisons of running cost of several image enhancement methods for three low-light images with different sizes (unit: second).}
	\begin{tabular}{|c|c|c|c|}
		\hline
		\multirow{2}{*}{Methods}  
		& \multicolumn{3}{|c|}{Image Size} \\ \cline{2-4}
		&$480 \times 640$&$640 \times 800$&$720 \times 1280$ 	 \\ \hline \hline
		DeHZ \cite{DeHZ} &$4.5540$ &$7.6023$ &$13.6565$ \\ \hline
		DePAMEF \cite{DePAMEF} &$12.9742$ &$22.8999$ &$39.3590$ \\ \hline
		BCP \cite{BCP} &$0.88181$ &$1.5076$ &$2.9533$ \\ \hline
		JIEP \cite{JIEP} &$3.3557$ &$5.6708$ &$8.8959$ \\ \hline
		FFM \cite{FFM} &$4.8740$ &$8.4554$ &$16.1714$  \\ \hline
		SRIE \cite{SRIE} &$5.3076$ &$8.8116$ &$16.1525$ \\ \hline
		LIME \cite{LIME} &$0.1592$ &$0.2113$ &$0.6426$ \\ \hline
		EETL \cite{EETL} &$1.5024$ &$1.5163$ &$1.5164$ \\ \hline
		Retinex-Net \cite{Retinex-Net} &$0.2675$ &$0.4558$ &$0.8891$ \\ \hline
		MBLLEN \cite{MBLLEN} &$0.4814$ &$2.9278$ &$5.5258$ \\ \hline
		Ours &$12.8165$ &$21.3407$ &$40.2050$  \\ \hline	
	\end{tabular} \label{table7}
\end{table}

\subsection{Experimental Results on Realistic Maritime Images}
\label{sub-sec:ERMRI}
Due to the distinctness between synthetic and realistic images, this subsection mainly focuses on low-light enhancement experiments on realistic images. Our proposed method will be compared with ten different imaging methods, i.e., DeHZ \cite{DeHZ}, DePAMEF \cite{DePAMEF}, BCP \cite{BCP}, JIEP \cite{JIEP}, FFM \cite{FFM}, SRIE \cite{SRIE}, LIME \cite{LIME}, EETL \cite{EETL}, Retinex-Net \cite{Retinex-Net}, and MBLLEN \cite{MBLLEN}. To reflect the imaging performance more intuitively, the low-light image enhancement results and their associated magnified views are shown in Figs. \ref{Fig10}-\ref{Fig12}.

From the visual comparisons, we find that DeHZ \cite{DeHZ} and Retinex-Net \cite{Retinex-Net} suffer from obvious color distortions and unnatural appearances, which cause the degradation of visual image quality. BCP \cite{BCP} easily produces white false light at the junction of bright and dark regions, especially in Fig. \ref{Fig12}. The enhancement results yielded by JIEP \cite{JIEP} and FFM \cite{FFM} have the risk of insufficient enhancement. LIME \cite{LIME} and our proposed method can obtain satisfactory enhancement results compared with the other competing imaging methods. However, our method can achieve better visual effects on the balance of suppression of unwanted random noise and preservation of fine structural details. Our superior performance can be further confirmed by the quantitative results NIQE and BMTQI. It can be found that our results can obtain excellent index evaluation values. The superior performance of our method benefits from the regularized illumination optimization and deep noise suppression. 
\subsection{Experimental Results on Realistic benchmark Images}
\label{sub-sec:ERRI}
To verify that our method can handle various low-light images robustly, we use three benchmark low-light images for experiments, and verify the superiority of our method by comparing with ten different imaging methods, i.e., DeHZ \cite{DeHZ},  DePAMEF \cite{DePAMEF}, BCP \cite{BCP}, JIEP \cite{JIEP}, FFM \cite{FFM}, SRIE \cite{SRIE}, LIME \cite{LIME}, EETL \cite{EETL}, Retinex-Net \cite{Retinex-Net}, and MBLLEN \cite{MBLLEN}. Fig. \ref{Fig13} shows the low-light image enhancement results and their associated magnified views generated by various methods.

By comparison, it can be clearly found that BCP \cite{BCP} and Retinex-Net \cite{Retinex-Net} have obvious color distortions and blocking artifacts. As can be seen from the magnified views of Image13 and Image14, DeHZ \cite{DeHZ}, LIME \cite{LIME}, and MBLLEN \cite{MBLLEN} fail to preserve the image details. The enhanced results yielded by EETL \cite{EETL} have a certain degree of color deviation. However, the enhancement results generated by JIEP, FFM, and SRIE. From the magnified view of Image15, JIEP \cite{JIEP}, FFM \cite{FFM}, and SRIE \cite{SRIE} have the risk of insufficient enhancement. However, Our method can not only enhance the details of dark regions, but also preserve the texture structure. To further prove the superiority of our method, we use two non-reference metrics (i.e., NIQE and BMTQI) to evaluate the enhanced image and organize the evaluation results in Table \ref{table6}. It can be seen that although our evaluation results on the single image fail to obtain optimal values, the average value of the evaluation results is optimal due to the robustness of our method.

\subsection{Comparisons of Running Time}
\label{sub-sec:RTC}
To analyze the computational time under different imaging conditions, we select three low-light images with sizes of $480 \times 640$, $640 \times 800$, and $720 \times 1280$ as test images. Our method will be compared with ten different image enhancement methods, i.e., DeHZ \cite{DeHZ},  DePAMEF \cite{DePAMEF}, BCP \cite{BCP}, JIEP \cite{JIEP}, FFM \cite{FFM}, SRIE \cite{SRIE}, LIME \cite{LIME}, EETL \cite{EETL}, Retinex-Net \cite{Retinex-Net}, and MBLLEN \cite{MBLLEN}, by calculating the running time for three experimental images with different sizes. The computational time of the competing image enhancement methods is summarized in Table \ref{table7}. LIME yields the lowest computational cost due to the fast variational method, but it sometimes suffers from slight artifacts in enhanced images. The relatively lower computational time could be generated using the deep learning-based image enhancement methods (i.e., EETL, Retinex-Net, and MBLLEM) since it is efficient to perform the well-trained networks. In contrast, our method takes a longer time to perform low-light image enhancement. However, our method is able to generate the superior enhancement performance in terms of both quantitative and qualitative image quality evaluations. Fortunately, the graphics processing unit (GPU) \cite{HuangIoT2020} has rapidly evolved into a cost-effective parallel computing platform, which has been successfully adopted to accelerate regularized variational model \cite{GuloJIP2019} and low-light image enhancement \cite{YangJIP2019,ParkIEEECE2017}. Thus, there will be a great incentive to accelerate our image enhancement method for real-time imaging applications in the GPU computing platform.
\section{Conclusion}
\label{sec:C}
In this work, we proposed to enhance low-light images by performing regularized illumination optimization and blind noise reduction. In particular, the hybrid regularized variational model was presented to perform structure-preserving illumination refinement. The final enhanced images were generated by combining the refined illumination and optimized reflection maps. The deep learning method was further introduced to eliminate the negative effect of unwanted noise on imaging performance. Owing to the regularized illumination optimization and deep noise suppression, our image enhancement method has the capacity of generating more natural-looking enhanced images under different low-light conditions. Comprehensive experiments on both synthetic and realistic maritime images have illustrated the effectiveness of our proposed method.
\ifCLASSOPTIONcaptionsoff
  \newpage
\fi

\end{document}